\documentclass[prl,superscriptaddress,preprint,showpacs,amsmath,amssymb]{revtex4-1}

\usepackage{graphicx}
\usepackage{indentfirst}
\usepackage{psfrag}
\usepackage{epsfig}
\usepackage{amsmath}
\usepackage{amssymb}
\usepackage{bm}

\def\be{\begin{eqnarray}}
\def\ee{\end{eqnarray}}

\begin{document}

%%%%%%%%%%%%%%%%%% title page information %%%%%%%%%%%%%%%%%%
%\title{Simple template for authors submitting to \textit{Optics Express}}
\title{Optical forces on cylinders near subwavelength slits illuminated by a photonic nanojet}
%\author{M. Scott Dineen and Jennifer Martin}
\author{F.J. Valdivia-Valero$^{1}$ and M. Nieto-Vesperinas$^{1,*}$}
%\address{Optics Express Office, Publications Department, Optical Society of America, \\ Washington, D.C., 20036}
\address{$^{1}$Instituto de Ciencia de Materiales de Madrid, C.S.I.C., Campus de Cantoblanco \\ 28049 Madrid, Spain}
%\email{opex@osa.org} %% email address is required
\email{$^{*}$mnieto@icmm.csic.es}


\begin{abstract}
We discuss optical forces exerted on particles, either dielectric or metallic,  near a subwavelength slit illuminated by a photonic nanojet. We compare those cases in which the  Mie resonances are or are not excited. The configurations on study are 2D, hence those particles are infinite cylinders and, in order to obtain extraordinary transmission,  the  illuminating beam is p-polarized.  We show the different effects of these particle resonances on the optical forces:  while whispering gallery modes under those illumination conditions  weaken the force strength, this latter is enhanced by localized plasmon excitation.  Also, illuminating the slit with a nanojet enhances the optical forces on the particle at the exit of the aperture by a factor between 3 and 10 compared with illumination of the slit with a Gausian beam. In addition, the pulling force that such a small resonant metallic  particle suffers on direct  illumination by a nanojet, can change by the presence of the slit, so that it may become repulsive at certain lateral positions of the particle.
\end{abstract}
%pacs{42.50.Wk, 87.80.Cc, 42.25.Kb, 05.40.-a}

%42.50.Wk	Mechanical effects of light on material media, microstructures and particles
%37.10.Jk	Atoms in optical lattices
%37.10.Vz	Mechanical effects of light on atoms, molecules, and ions
%87.80.Cc	Optical trapping
%42.25.Kb	Coherence
%05.40.-a	Fluctuation phenomena, random processes, noise, and Brownian motion

\maketitle

%\ocis{(000.0000) General.} % REPLACE WITH CORRECT OCIS CODES FOR YOUR ARTICLE
%\ocis{(050.1940) Diffraction; (050.1220) Apertures;  (350.4855) Optical tweezers or
%optical manipulation; (230.5750) Resonators; (240.6680) Surface plasmons.}

%%%%%%%%%%%%%%%%%%%%%%% References %%%%%%%%%%%%%%%%%%%%%%%%%

%%%%%%%%%%%%%%%%%%%%%%%%%%  body  %%%%%%%%%%%%%%%%%%%%%%%%%%
\section{Introduction}

Studies on optical forces on micro and nano-objects, both in their applications to trapping \cite{Ashkin1986, Chu1990,  Yin1995, Liu1996, Okamoto1999, Chaumet2000, NietoVesperinas2004, Dholakia2008, Juan2009, Juan2011} as well as to optical binding \cite{Burns1989, Chaumet2001, Zemanek2010}, show the sensitivity of these techniques to the thermal action on the kinetics of these systems. This involves to increase both the numerical aperture NA and the power of the illuminating beam \cite{Juan2009} to control the experiments; the former procedure  has obvious limitations for the manipulation of biological specimens.

The effectiveness of optical trapping increases \cite{Juan2009, ValdiviaValero2012_6} by illuminating particles through subwavelength apertures in supertransmission \cite{GarciaVidal2010, Ebbesen1998, Garcia1979, GarciaMartin1997, NFO, Ripoll, Degiron2004, Lezec2004, Alu2006, Lezec2002, GarciaVidal2005, DiGennaro2010}, namely on excitation of the aperture morphology dependent resonances (MDR). This allows lower illumination powers  and its performance is greatly improved \cite{ValdiviaValero2012_6} when also  the Mie resonances of the particles are excited, i.e. their whispering gallery modes (WGM) \cite{AriasGonzalez2000, AriasGonzalez2001, Astratov2004, Chen2006, Deng2004, Boriskina2006} or localized surface plasmons (LSP) \cite{Maier2005, Sburlan2006, Maier2007, Pelton2008}. This enhances the aperture transmittance and localization of light \cite{ValdiviaValero2011_2, ValdiviaValero2010_1, ValdiviaValero2011_3}.

Nevertheless, illuminating with photonic nanojets (PNJ) \cite{Taflove2004, Taflove2005, Taflove2006, Taflove2009} constitutes an alternative means to enhance that transmission and localization of light through apertures \cite{ValdiviaValero2011_4}. PNJs have subwavelength spatial resolution \cite{Taflove2004, Itagi2005, Heifetz2007, Taflove2005} and hence are of great interest for microscopy and detection at the nanoscale \cite{Heifetz2006}. Since however they are nonresonant focusing effects, their appearence is not so narrowly constrained by the constitutive parameters and morphology of the particles as Mie resonances are.

As far as we know, in contrast with the extensive study on nanojets, there is only one report concerning  optical forces on a metallic particle close to a PNJ \cite{Cui2008}. We thus present here an analysis on the effect that the presence of subwavelength apertures has on these forces;  a subject that combines the phenomenon of extraordinary transmission with the nanojet focusing and localization of optical energy, and which  has never been addressed. The nanoparticle is either dielectric or metallic, and its MDRs may be excited.

We shall employ 2D configurations, so that the particles are cylinders with cross section in the plane of calculations, and the apertures are slits. It is well-known that this accounts for the main features of the phenomenon providing one does not look for depolarization effects \cite{ValdiviaValero2012_6, Ho1994, AriasGonzalez2002, Chaumet2000_2}and it is certainly the case for PNJs \cite{Taflove2004, Taflove2005, Taflove2006, Taflove2009} . Therefore in order to obtain slit supertransmission, the incident wave is p-polarized.

We will employ the Maxwell stress tensor (MST) \cite{Okamoto1999, NietoVesperinas2004, AriasGonzalez2002, Chaumet2000_2, Chaumet2009, NietoVesperinas2010, Blanco2007, Albaladejo2009, ValdiviaValero2012_6}. The fields are calculated by a finite element method  \cite{ValdiviaValero2012_6}. Section 2 accounts for these computations as well as for those of  the energies and electromagnetic forces, whose details were given in our previous works \cite{ValdiviaValero2010_1, ValdiviaValero2011_4, ValdiviaValero2012_6}.

Then we shall analyze in Sections 3 and 4 the effects of exciting a  WGM in a $Si$ cylinder and a LSP in an $Ag$ one, respectively, on the optical forces exerted on them. This is carried out on comparing the effect of PNJ illumination with and without a subwavelength slit. In this way  we shall show that whereas  a metallic particle as small as that addressed  here, resonantly illuminated by a nanojet, suffers an attractive vertical force, (this is in contrast with previous results \cite{Cui2008} concerning larger metallic particles),  which makes the  nanojet forming cylinder a  photonic tractor,  the additional presence of a subwavelength aperture may change the sign of this force thus transforming it to repulsive, depending on the lateral position of the cylinder.

In addition, if we previously showed \cite{ValdiviaValero2012_6} that the presence of the slit enhances the gradient forces by two orders of magnitude compared with the direct illumination of the probe particle  by a Gaussian beam  of a conventional optical tweezer, we now observe that illuminating the slit with a  nanojet increases the magnitude of the optical forces on the particle by a factor between 3 and 10 compared with illumination of the slit with a Gaussian beam.

%Standard \LaTeX{} or AMS\TeX{} environments should be used to place tables,
%figures, and math. Examples are given below.

%\begin{verbatim}
%\begin{figure}[htbp]
%\centering\includegraphics[width=14cm]{} \caption{Sample caption
%(Ref. \cite{Oron03}, Fig. 2).}
%\end{figure}

%\begin{equation}
%H = \frac{1}{2m}(p_x^2 + p_y^2) + \frac{1}{2} M{\Omega}^2
%     (x^2 + y^2) + \omega (x p_y - y p_x).
%\end{equation}
%\end{verbatim}

\section{Numerical calculations}
The  calculation of the electromagnetic fields interacting with particles placed at each side of an aperture in a metal slab is now discussed.  2D numerical simulations are done by means of a finite element method (FE) (FEMLAB 3.0a of COMSOL, \mbox{http://www.comsol.com}). Aside polarization effects,  the main features of the physical process of coupling, resonance excitation and nanojet focalization are analogous to those in 3D; this was already analyzed for nanojets  in \cite{Taflove2004, Taflove2005, Taflove2006, Taflove2009}.  Technical details on  the calculations have been given in \cite{GarciaPomar2005, ValdiviaValero2010_1, ValdiviaValero2012_6}.

Although other materials can be chosen, $Si$ and $Ag$ have been used to simulate the probe particles, performing the analysis on the grounds of their rich Mie resonance spectra in the studied regimes, (near IR and near UV, respectively). The dielectric material for the nanojet-focusing particle is $SiO_{2}-glass$ \cite{Johnson1972, Palik1998}. The reason of working with either dielectric or metallic particles is to see the effects of their Mie resonances on the induced electromagnetic forces, taking into account the influence of the presence of a subwavelength slit illuminated by the nanojet. As in previous works, and because of the same reasons \cite{Rigneault2005, ValdiviaValero2011_4, ValdiviaValero2012_6},  the latter is  considered in a thick $Al$ slab.  Noble metals like $Au$ or $Ag$ may be used for thinner slabs.

\begin{figure}[htbp]
\centering
\includegraphics[width=8cm]{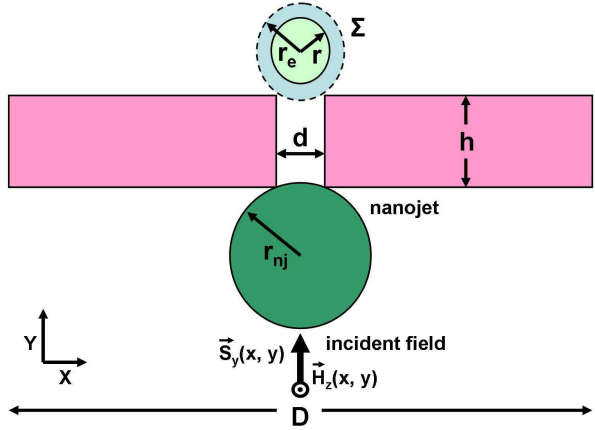}
\caption{Schematic illustration of the geometry for the calculation of the slit transmittance and optical force calculations: An incident p-polarized plane wave with a Gaussian profile, (amplitude $A = (1/\pi)mW/\mu m^2$, width of its intensity $\sigma = r_{nj}$), magnetic vector ${\bf H}_{z}$ and time-averaged Poynting vector $<{\bf S}_{y}>$ impinges a $SiO_{2}-glass$ cylinder of radius $r_{nj}$ which focuses a nanojet near its back surface, whose intensity is calculated at the point where $|<{\bf S}_{y}>|$ is maximum. This nanojet illuminates, either directly or by interposing a subwavelength aperture of widht $d$, a cylinder of radius {\it r}. The aperture is  practiced in an {\it Al} slab of width {\it D} and thickness {\it h}. The width {\it D} is also that of the horizontal side of the simulation window, whose boundary conditions are set to {\lq\lq}low reflection", except in those segments which coincide with the exterior limits of the slab, where  {\lq\lq}perfect conductor" conditions are set. The intensity focused by the nanojet on the cylinder of radius {\it r}, either directly or  on transmission through the slit,  is evaluated as follows:  For dielectric ($Si$) particles, the time averaged energy flow norm $|<{\bf S}_{y}>|$ is integrated inside the cylinder circle cross section of radius $r$. For metallic ($Ag$) cylinders, one determines $|<{\bf S}_{y}>|$  integrated in an annulus of exterior and interior radii: $r_{e}$ and  $r$, respectively. As a beam of unit amplitude is always launched on the system, all transmission graphics presented in this work are normalized to this value. The circumference $\Sigma$ of radius $r_{e}$, is also used to calculate the electromagnetic forces on the probe particle (see also \cite{ValdiviaValero2012_6}), no matter whether the cylinder is dielectric or metallic. When there is no upper probe cylinder, the transmittance is obtained on integration of  $|<{\bf S}_{y}(x, y)>|$ in a   circle in vacuum which  coincides with  that of the probe cylinder cross section.}
\end{figure}

Following the scheme shown in Fig. 1, an incident beam, linearly p-polarized, namely with its magnetic vector ${\bf H}_{z}$ perpendicular to the XY-plane is launched upwards, propagating in the Y direction. This choice is because p-polarization, in contrast with s-polarization, produces   homogeneous eigenmodes, i. e. those which lead to extraordinary transmission of the subwavelength slit \cite{ValdiviaValero2010_1, ValdiviaValero2011_4, Jackson1999, Porto1999,  Garcia2007}. The light beam incides  on a dielectric cylinder of radius $r_{nj}$, which concentrates a nanojet near its back surface. The nanojet illuminates either the probe cylinder of radius $r$ when  the slab in Fig. 1 does not exist;  or,  when the latter is present, this illumination is done on transmission of the nanojet through the slit, as seen  in Fig. 1.  The nomenclature followed to classify both  localized surface plasmons (LSP) and whispering gallery modes (WGM) of the probe cylinders, as well as the  resonances of the slit is: subscripts \emph{(i, j)}, \emph{i} and \emph{j} standing for their angular \emph{i-th} and radial \emph{j-th} orders, respectively. In the case of the supertransmission resonances of the slit alone the subscripts \emph{(u, v)} will be used, \emph{u} and \emph{v} standing for their longitudinal \emph{u-th} and transversal \emph{v-th} orders.

In all cases the beam profile at frequency $\omega$ is Gaussian, with magnetic  field distribution at its focal plane: ${\bf H_z(x, y)}=|{\bf H}_{z0}|exp(-x^{2}/2\sigma^{2})exp(i((2\pi/\lambda)y-\omega t))$, $|{\bf H}_{z0}|$ denoting the modulus amplitude, $2^{1/2}\sigma$ being the half width at half maximum (HWHM) of the beam, and $\lambda$ representing its wavelength. In this way the beam has intensity $(1/\pi)mW/\mu m^2$ and $HWHM=2r_{nj}$. The geometrical parameters of the cylinders and of the  slit have been adjusted such that at the given illuminating wavelength $\lambda$, there is focalization of nanojets as well as excitation of morphology-dependent resonances of both the slit and the probe particle.

The physical quantities shown in the images of the configurations under study are the magnetic field ${\bf H}_{z}(x, y)$, the electric field ${\bf E}(x, y)$ and the time-averaged energy flow $<{\bf S}(x, y)>$. The curves presenting the nanojet intensity focused near the back surface of the $SiO_{2}-glass$ cylinder, are obtained at the point $(x, y)$ where  $|<{\bf S}(x, y)>|$ is maximum.

The light concentration either in or on the probe cylinder according to whether this is dielectric or metallic, is evaluated by integrating $|<{\bf S}(x, y)>|$ in a circle which coincides with  the circle cross section of the probe cylinder of radius $r$ when this latter is dielectric, or in an annulus surrounding it of radii  $r$ and  $r_{jm}$ when the probe cylinder is metallic,  see Fig. 1.  This stems from the fact that  if the particle is dielectric  the intensity transmitted by the slit, that couples to the particle WGM, is concentrated inside the cylinder;  whereas when this nanoparticle is metallic, this transmitted intensity, coupled to a LSP, remains on the particle surface.  On the other hand, the transmittance of the slab when the probe particle  is not present is calculated on integration of $|<{\bf S}(x, y)>|$ in a circle of radius $r$ in void space which coincides with that of the probe cylinder cross section.  In all cases these intensities are normalized to the maximum intensity of the incident Gaussian beam $|<{\bf S}_{max}>|= (1/\pi)mW/\mu m^2$.

The time-averaged force on the probe cylinder is calculated by employing  the  Maxwell stress tensor (MST) \cite{Jackson1999}:

%\begin{equation}
\begin{align}
{<\bf  F}_{em}>= &\int\int_\sum[\epsilon/2 Re\{({\bf E}\cdot
{\bf n}){\bf E^\ast}\}-\epsilon/4 ({\bf E}\cdot {\bf
E^\ast}) {\bf n}+\mu/2  Re\{({\bf H}\cdot {\bf n}){\bf
H^\ast}\} \nonumber \\
&-\mu/4 ({\bf H}\cdot {\bf H^\ast}) {\bf n}] ds, \label{eq:MET}
\end{align}
%\end{equation}

{\noindent where the surface of integration $\Sigma$ surrounds the particle as seen in Fig. 1 and ${\bf n}$ represents}  the outward unit normal. In our 2D geometry, $\Sigma$ is the circumference of radius $r_e$, (see Fig. 1). $\epsilon=\mu=1$.

In Eq. (1), ${\bf E}, {\bf H}$ and ${\bf E^\ast}, {\bf H^\ast}$ are the values of the fields and their complex conjugates, $\epsilon$ and $\mu$ denotes the electric permittivity and magnetic permeability of the surrounding medium, which in  this work will be assumed to be vacuum.

The calculation with FEMLAB of the complex values ${\bf E(r)}$ and ${\bf H(r)}$ and of the real physical fields: ${\bf E^R}({\bf r}, t) = Re[{\bf E(r)}\exp(-i\omega t)]$ and ${\bf H^R}({\bf r}, t) = Re[{\bf
H(r)}\exp(-i\omega t)]$, are not straightforward. The details of the procedure have been given in \cite{ValdiviaValero2012_6}

%\section{Response of in extraordinary transmission of a slit - cylinder system}

%\subsection{Effects on supertransmission due to a passive cylinder located beyond or before the slit}

%\subsection{Electromagnetic forces on a passive cylinder located beyond or before the slit}

\section{Extraordinary transmission of a slit in presence of a nanojet. Excitation of a whispering gallery mode}

\subsection{Excitation of a WGM in a cylinder by nanojet focalization.  Effects of coupling by supertransmission}

In order to interpret the optical forces on the probe particles, we first discuss the fields resulting from the interaction. We begin dealing  with the situation in which the slit is illuminated by a nanojet  in the infrared region. The probe cylinder on the slit exit being dielectric. This makes it possible two processes of supertransmission enhancement: one due to the nanojet focused near the back surface of the dielectric cylinder of radius $r_{nj}$ at the entrance of the aperture, and that resulting from the  coupling between the light exiting the slit and the WGM excited in the probe cylinder of radius $r$, (see Fig. 1).

\begin{figure}[htbp]
\begin{minipage}{.49\linewidth}
\centering
\includegraphics[width=6cm]{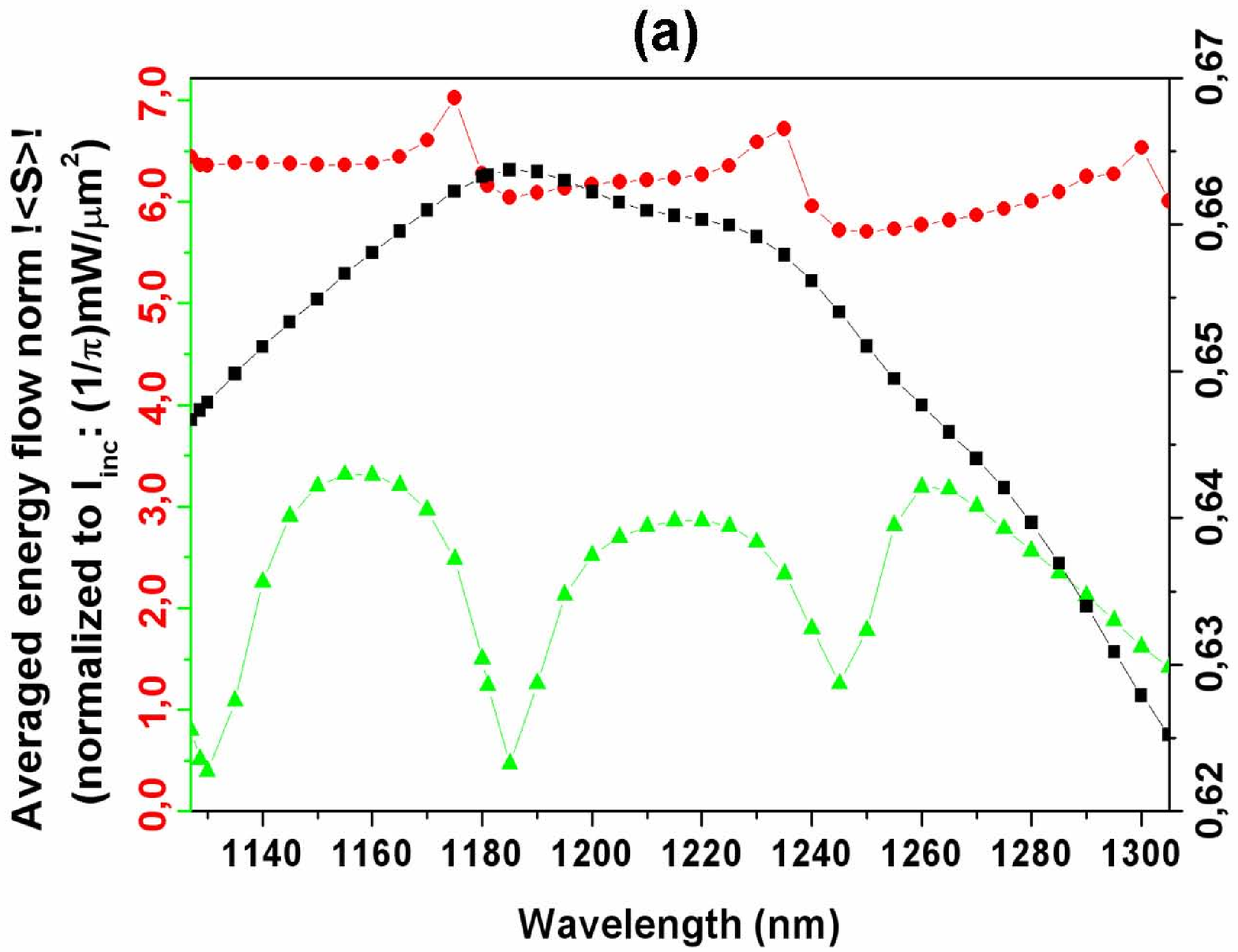}
\end{minipage}
\begin{minipage}{.49\linewidth}
\centering
\includegraphics[width=6cm]{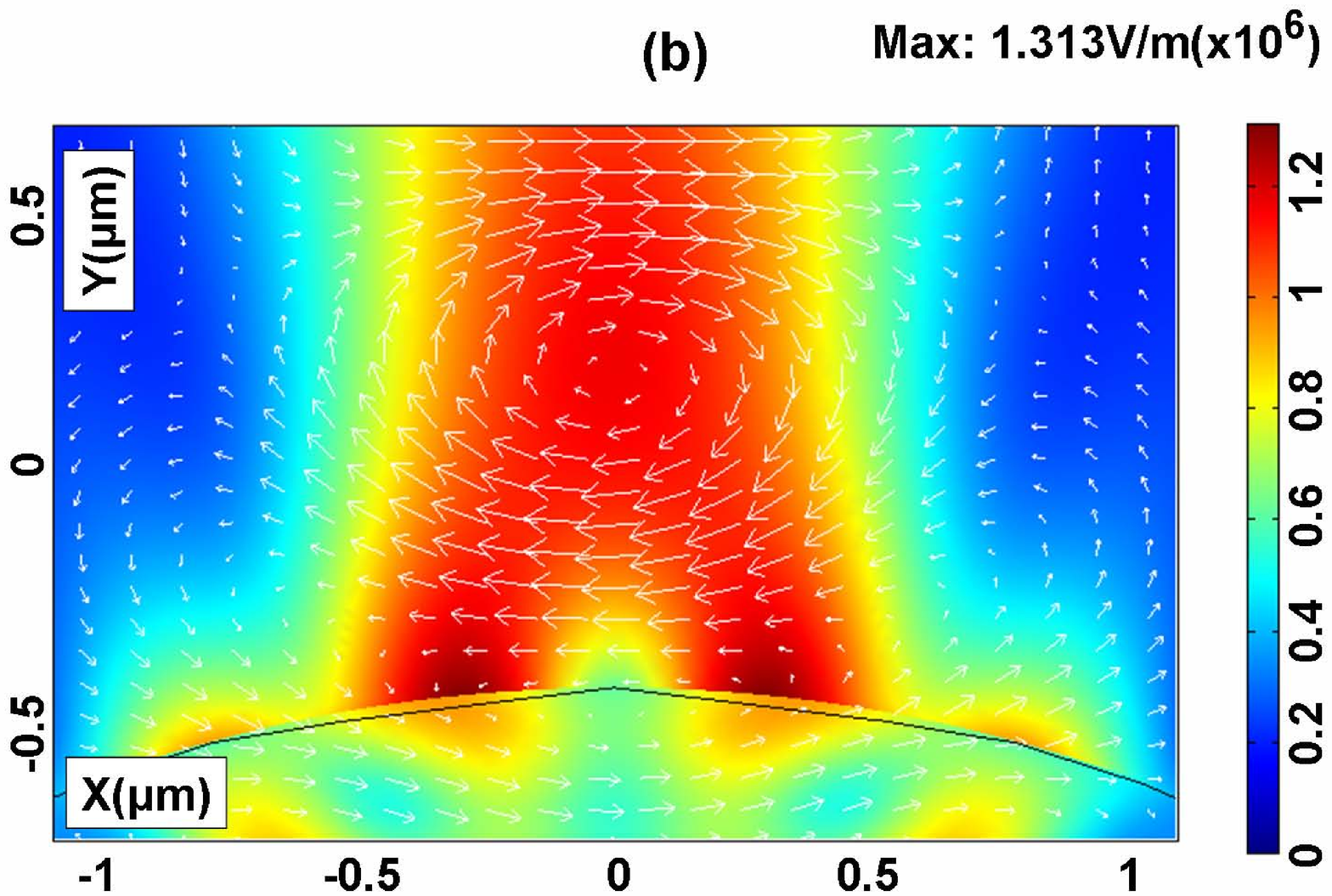}
\end{minipage}
\begin{minipage}{.98\linewidth}
\centering
\includegraphics[width=6cm]{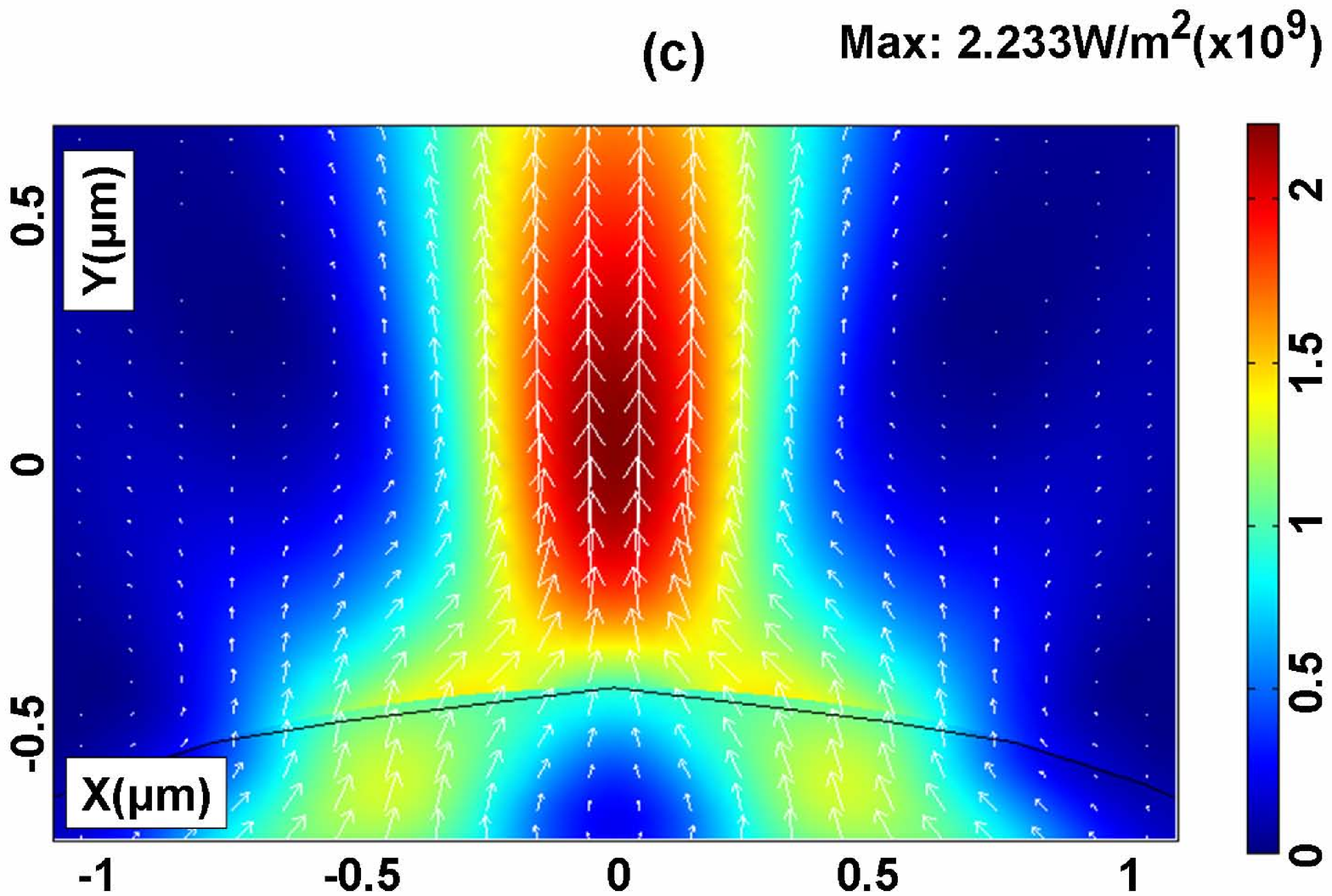}
\end{minipage}
\caption{(a) Time-averaged energy flow norm ${\bf |<S(r)>|}$, transmitted by the slit against illumination wavelength of:  a slit alone (width $d = 428.8nm$) practiced in an {\it Al} slab (width $D = 19.920\mu m$, thickness $h = 857.6nm$) (black squared curve, values in right vertical axis); and of the same slit illuminated by a nanojet focused near the back surface of a $SiO_{2}-glass$ cylinder of radius $r_{nj} = 3\mu m$ placed at the entrance of this slit (green triangle curve, values in left vertical axis). The red circle curve (values in left vertical axis) stands for ${\bf |<S(r)>|}$ concentrated by the nanojet at its maximum intensity point when the slab is absent. The calculations are done as explained in Fig. 1. (b) Detail of the electric field ${\bf |E(r)|}$ magnitude transmission (colors) and vector (arrows) of the nanojet focused near the back surface of the $SiO_{2}-glass$ cylinder at $\lambda = 1175nm$, without slit nor $Si$ probe particle. This corresponds to a peak of the red circle line of (a). (c) The same  detail as in (b) but of  the time-averaged energy flow ${\bf <S(r)>}$, (norm in colors and vector in arrows) .}
\end{figure}

We first make a study of the fields and energy without the probe $Si$.  Fig. 2(a) shows the black squared and green triangled curves which stand for the   average energy flow, either when the Gaussian beam directly illuminates the $Al$ slab without the presence of the nanojet, (peak at $\lambda = 1175nm$), or when the incident light is focused as a nanojet near the back surface of a $SiO_{2}-glass$ cylinder of radius $r_{nj}=3\mu m$ placed at the slit entrance. The intensity at that point of the nanojet where it is maximum, without slab, is plotted as the red circle curve. The presence of the nanojet enhances the transmittance peak of the slit by a factor above 4  (at $\lambda = 1155nm$), but erases the supertransmitted energy peak profile of the slit alone (compare the green triangle and the black square lines), manifesting the non resonant nature of the PNJ focusing process (see the red circle line). A detail of the nanojet alone, i.e. without slit nor $Si$ cylinder, near the back surface of the cylinder is displayed both by its  electric field ${\bf E(r)}$ distribution [see Fig. 2(b)] and by that of its energy flow ${\bf <S(r)>}$ [Fig. 2(c)], both maps clearly show the focalization process and the interesting vortex of electric field lines about the nanojet focus.

\begin{figure}[htbp]
\begin{minipage}{.49\linewidth}
\centering
\includegraphics[width=6cm]{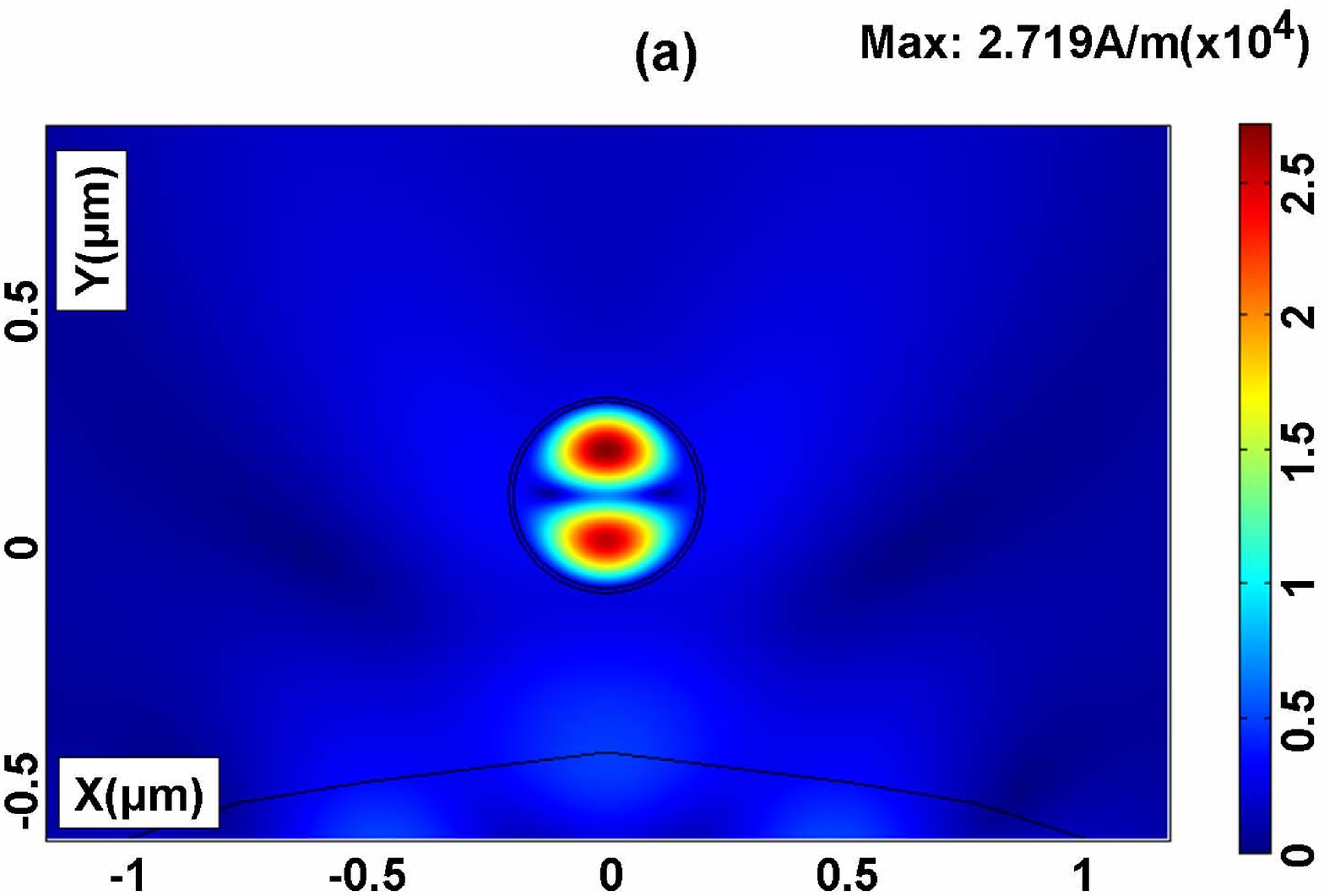}
\end{minipage}
\begin{minipage}{.49\linewidth}
\centering
\includegraphics[width=6cm]{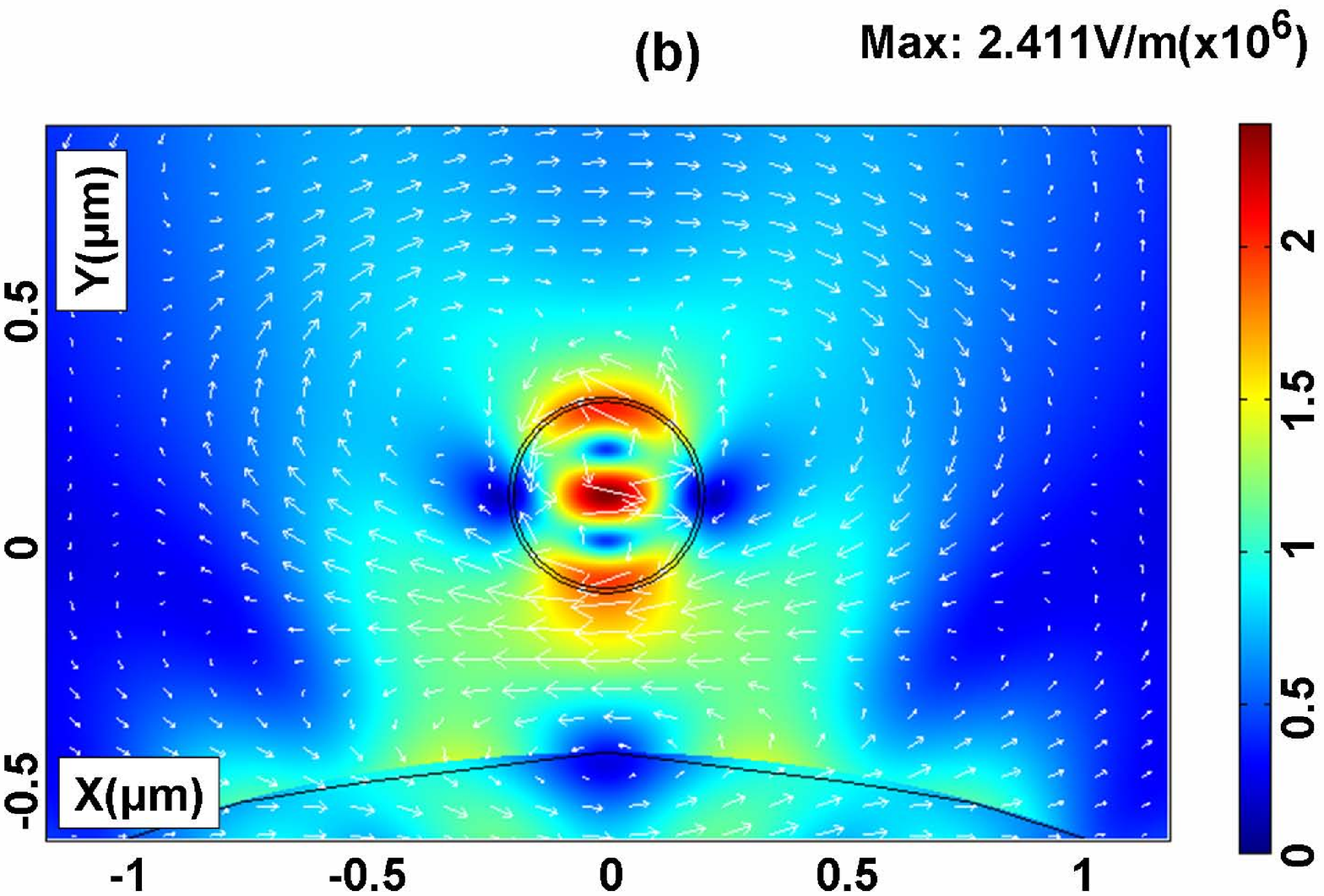}
\end{minipage}
\begin{minipage}{.98\linewidth}
\centering
\includegraphics[width=6cm]{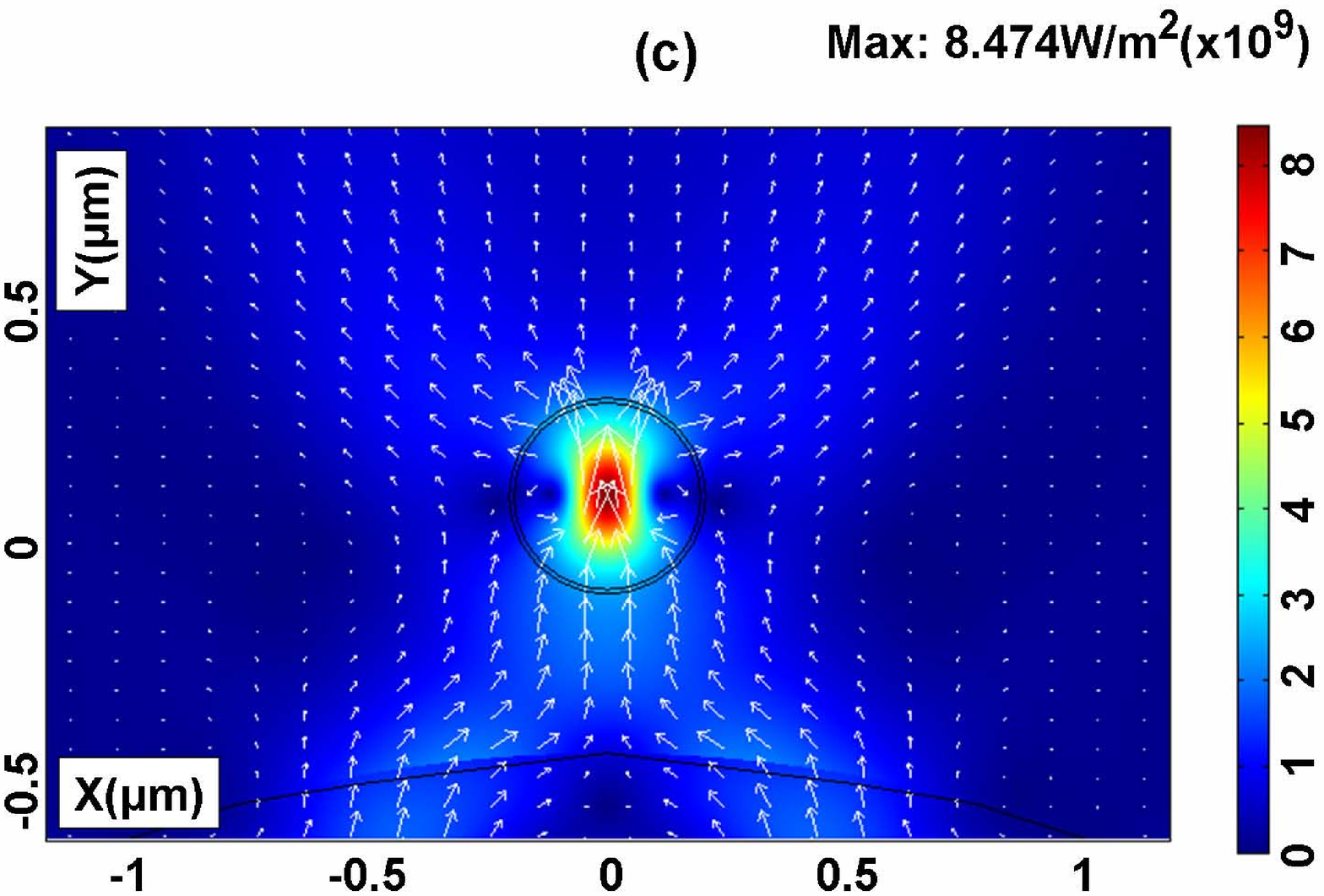}
\end{minipage}
\caption{(a) Detail of the map of the magnetic field norm ${\bf |H(r)|}$, ${\bf r}=(x,y)$,   in a resonant {\it Si} cylinder of radius $r = 200nm$ illuminated by the nanojet focused at {\it 350nm} near the back surface of the $SiO_{2}-glass$ cylinder of radius $r_{nj}=3\mu m$.  The $Si$ cylinder is placed at the maximum intensity point of the nanojet. There is energy concentration as a $WGH_{2,1}$ mode of the $Si$ cylinder. (b) Detail of the electric field ${\bf E(r)}$ magnitude and vector, (colors and arrows, respectively).  (c) Detail of time-averaged energy flow ${\bf <S(r)>}$, (norm in colors and vector in arrows). The conditions of illumination in these images are those of Figs. 2(b) - (c) and at the same wavelength.}
\end{figure}

Next, the $Si$ probe cylinder is placed above the $SiO_{2}-glass$ without the presence of the slit. The light in the $Si$ cylinder now concentrates in the form of a $WGH_{2,1}$ resonance, as shown by the dipolar map of the magnetic field norm ${\bf |H(r)|}$ in Fig. 3(a), ${\bf r}=(x,y)$. Consequently, the electric field ${\bf E(r)}$ distribution acquires a vortex pattern near the probe cylinder [cf. Fig. 3(b)] and, under  this incident p-polarization, the light intensity from the nanojet is strongly localized within this $Si$  cylinder [see Fig. 3(c)].

\begin{figure}[htbp]
\begin{minipage}{.49\linewidth}
\centering
\includegraphics[width=6cm]{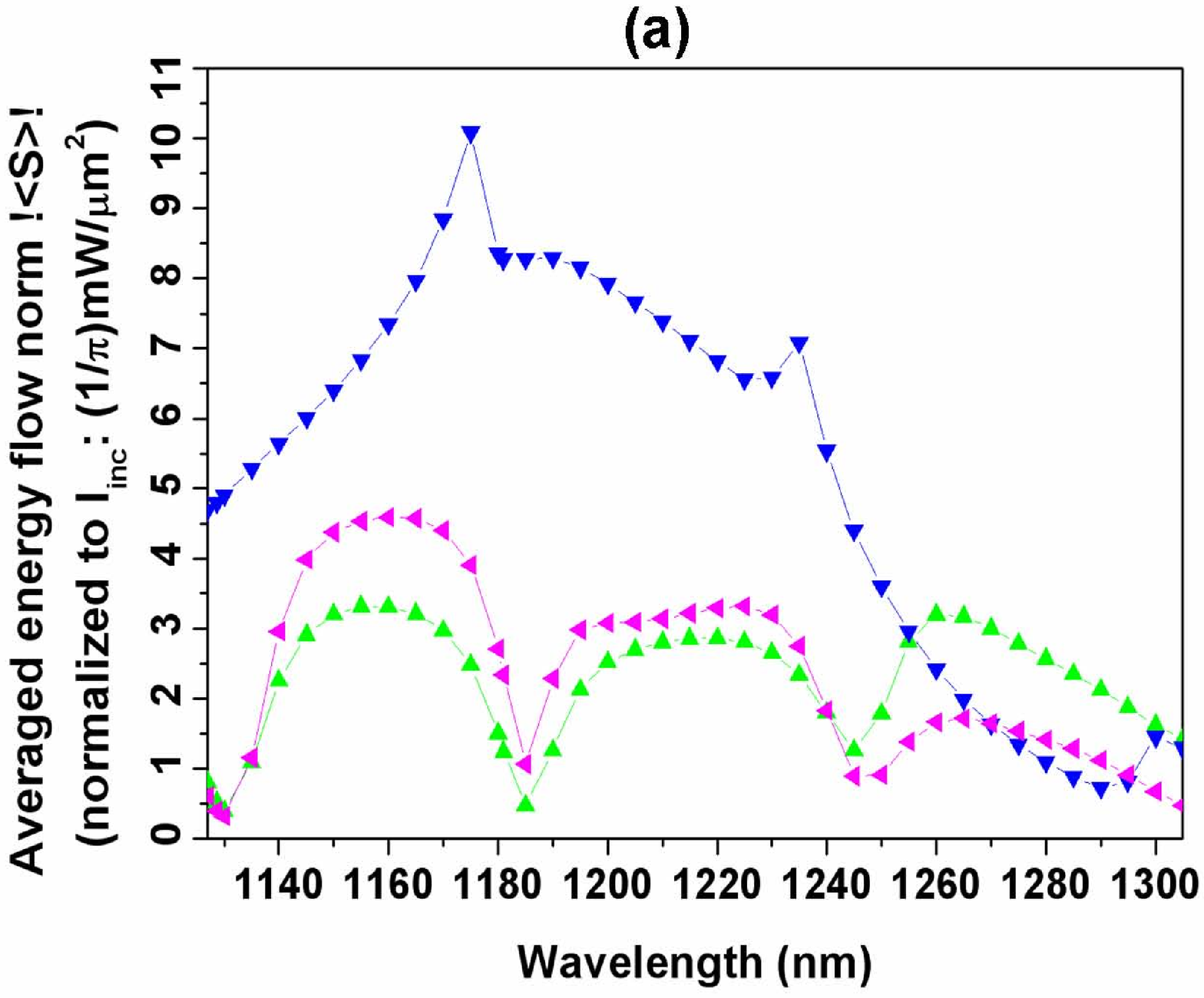}
\end{minipage}
\begin{minipage}{.49\linewidth}
\centering
\includegraphics[width=6cm]{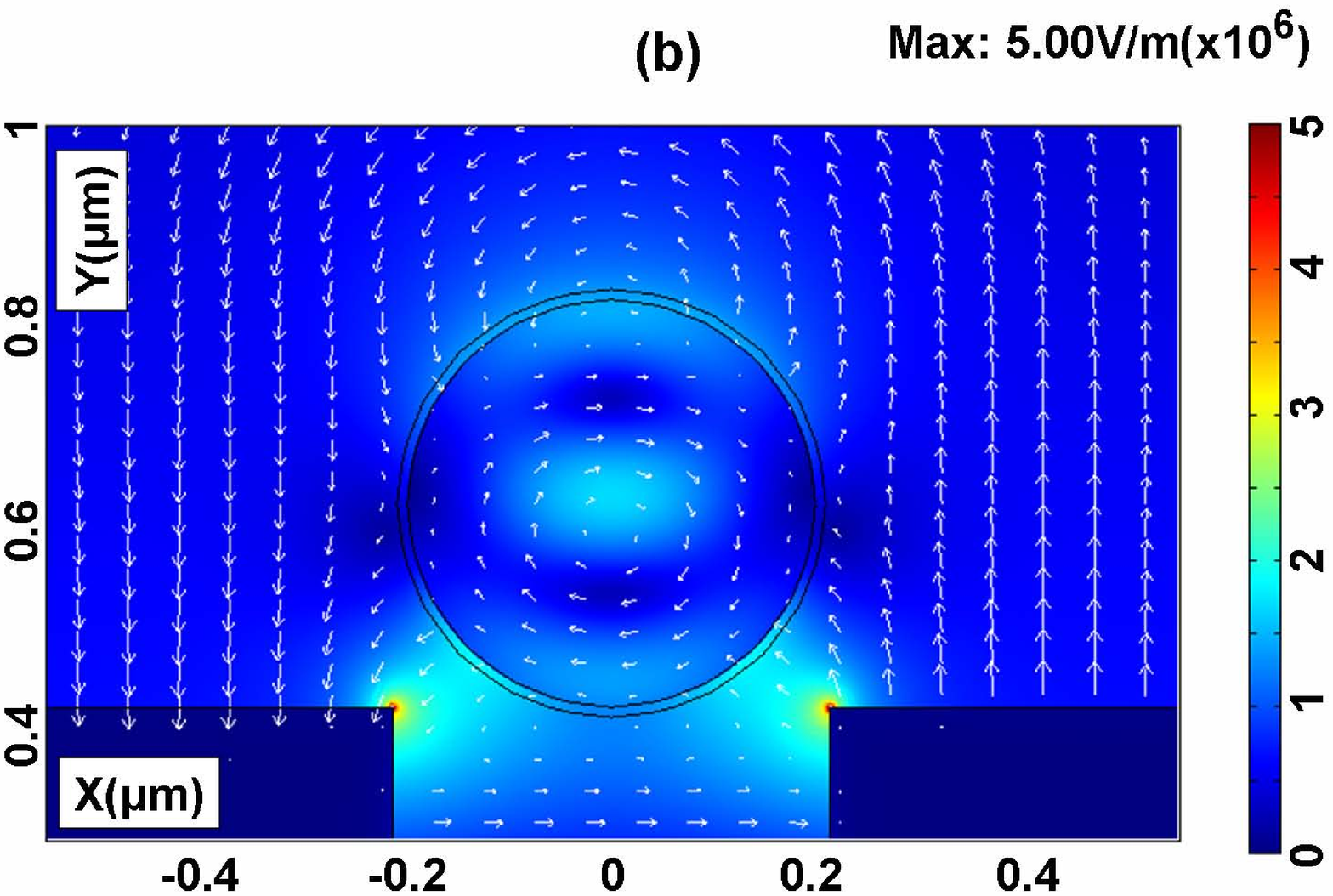}
\end{minipage}
\begin{minipage}{.98\linewidth}
\centering
\includegraphics[width=6cm]{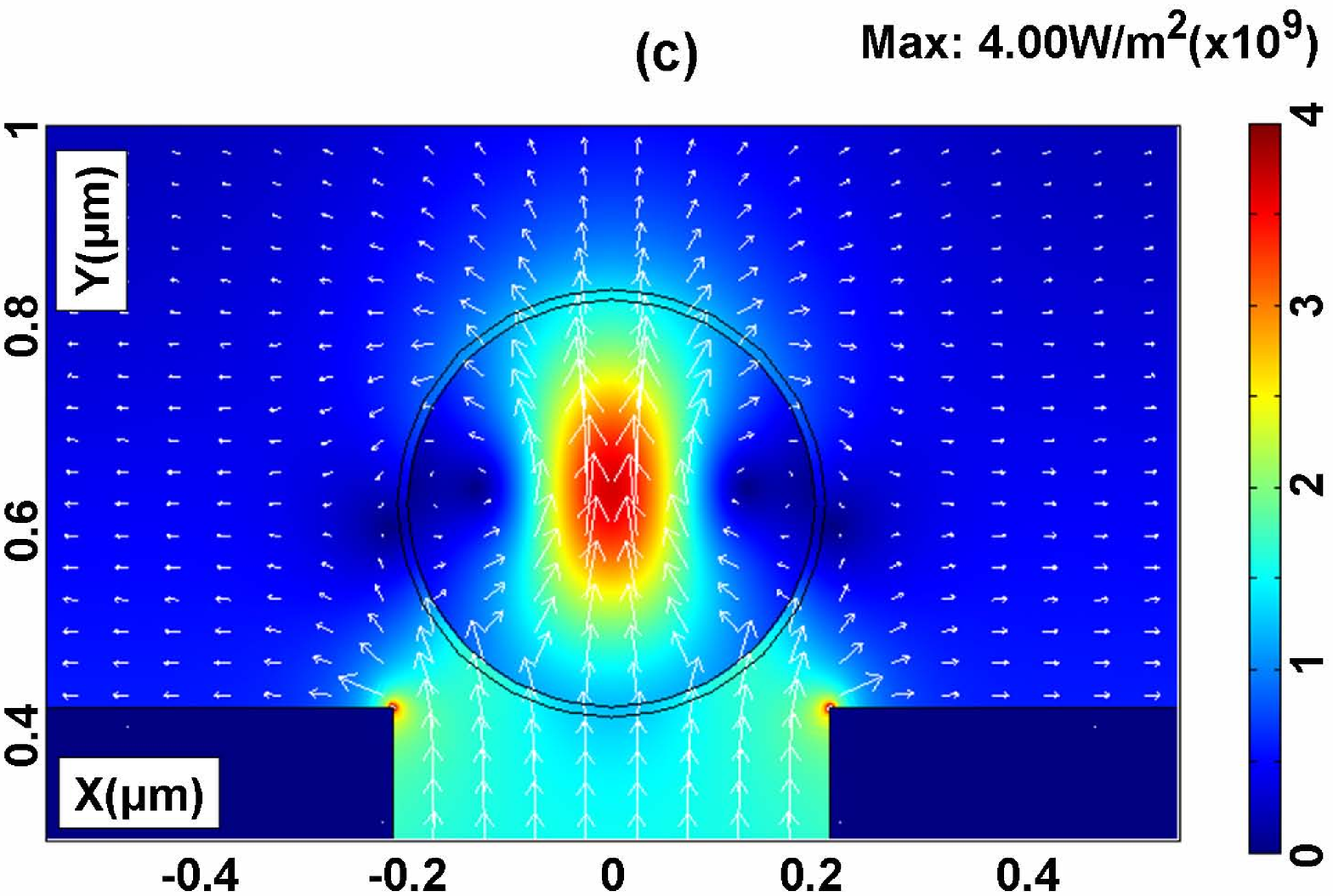}
\end{minipage}
\caption{(a) Transmitted time-averaged energy flow norm ${\bf |<S(r)>|}$ versus illumination wavelength of the following configurations: A nanojet focused by the $SiO_{2}-glass$ cylinder  of radius $r_{nj}=3\mu m$, tangent to the entrance plane of the slit practiced in the {\it Al} slab studied in Fig. 2(a), (green curve with up-triangles). A probe {\it Si} cylinder is added, tangent to the exit plane of the slit illuminated by the nanojet (pink curve with rotated triangles). The blue curve with down-triangles  stands for ${\bf |<S(r)>|}$ concentrated in the resonant {\it Si} cylinder,  placed above the nanojet-focusing $SiO_{2}-glass$ one,  when there is no slab between them.  All these curves are obtained as explained in Fig. 1.   (b) Detail of the electric field ${\bf E(r)}$ magnitude and vector (colors and arrows) in the {\it Si} cylinder at the exit plane of the  slit illuminated by the nanojet, (see in (a) the pink curve with rotated triangles). The  illumination of the  nanojet forming  large $SiO_{2}-glass$ cylinder below the slit is with a Gaussian beam at $\lambda = 1160nm$.  This cylinder is not shown in the figure. (c) The same as in (b) for ${\bf <S(r)>}$, norm in colors and vector in arrows.}
\end{figure}

Figure 4(a) shows the   average energy flow for all the configurations studied in this paper. The intensity of the nanojet localized in the $Si$ cylinder of radius $r_{nj}=3\mu m$, without slit,  is plotted in the blue curve with triangles down. The peaks of this curve coincide with those of the intensity peak point of the nanojet alone, [see the red curve with circles in Fig. 2(a)], but now enhanced by the resonant concentration due to the $WGH_{2,1}$ resonance at $\lambda = 1175nm$. As seen, this nanojet also enhances the energy asociated to the cylinder resonance by a factor of 10, [cf. the blue curve with triangles down in Fig. 4(a)], on comparing with the case of the resonant $Si$ cylinder directly illuminated by the Gaussian beam without nanojet nor slit, (whose intensity integrated in the $Si$ cylinder cross-section has a maximum at $\lambda = 1195nm$; not shown here for brevity).

Concerning the nanojet focused in the $Al$ slit with the $Si$ cylinder at its exit, the transmitted intensity inside that cylinder against wavelength takes the form of the pink curve with rotated triangles of Fig. 4(a). This follows the trend of the intensity in that cylinder alone, (not shown here), modulated by that of the nanojet-slit system, as a result of the effect of nanojet focalization plus WGM resonance excitation.  This shows a modest increase of transmitted intensity inside the $Si$ particle (cf. the pink curve with rotated triangles) comparing with that  of the nanojet large cylinder plus slit, without the $Si$ particle, (cf. the green curve with triangles up).  Nevertheless, the enhancement factor is now above 4 (at $\lambda = 1160nm$) with respect to the transmitted intensity from the slit alone, [cf. black curve with squares in Fig. 2(a)].

A detail of the spatial distributions of ${\bf E}$ and $<{\bf S}>$  at the optimum wavelength for largest transmitted intensity of the  system constituted by the large $SiO_{2}-glass$ cylinder,  the subwavelength slit and the resonant $Si$ probe cylinder, can be seen in Figs. 4(b) and 4(c), respectively, showing the vortex of the electric vector in the $Si$ cylinder around the point of maximum intensity concentration. Also, the large charge localization in the corners of the slit is clearly observed. This will be seen to have important consequences for the optical forces on the $Si$  cylinder.

\subsection{Electromagnetic forces on a dielectric cylinder with a whispering gallery mode excited by a nanojet,  either directly or on transmission  through a subwavelength slit.}

Next, we study the force that the electromagnetic fields of the nanojet transmitted through the slit exert on the probe $Si$ cylinder  placed at the slit exit.

\begin{figure}[htbp]
\begin{minipage}{.49\linewidth}
\centering
\includegraphics[width=6cm]{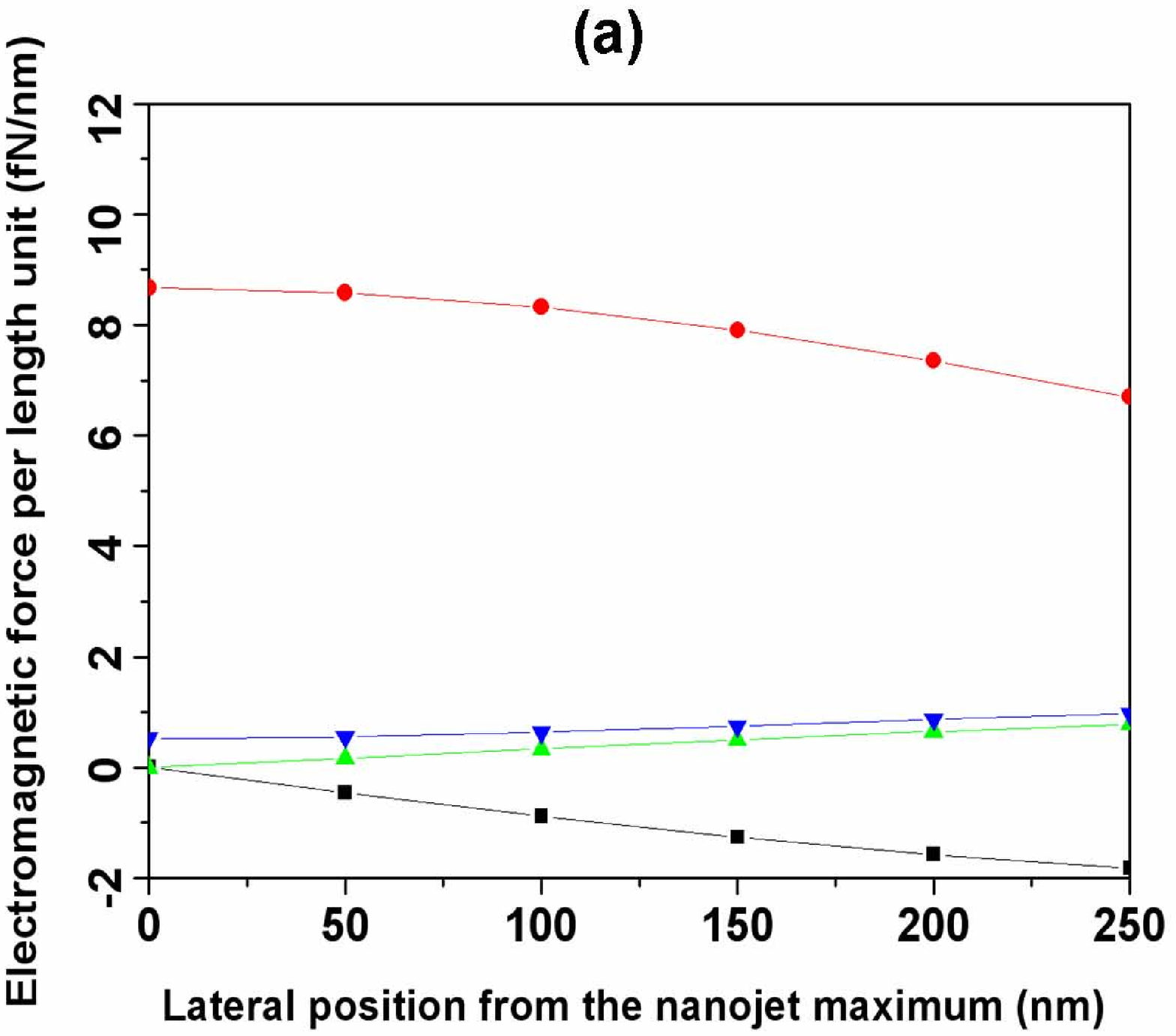}
\end{minipage}
\begin{minipage}{.49\linewidth}
\centering
\includegraphics[width=6cm]{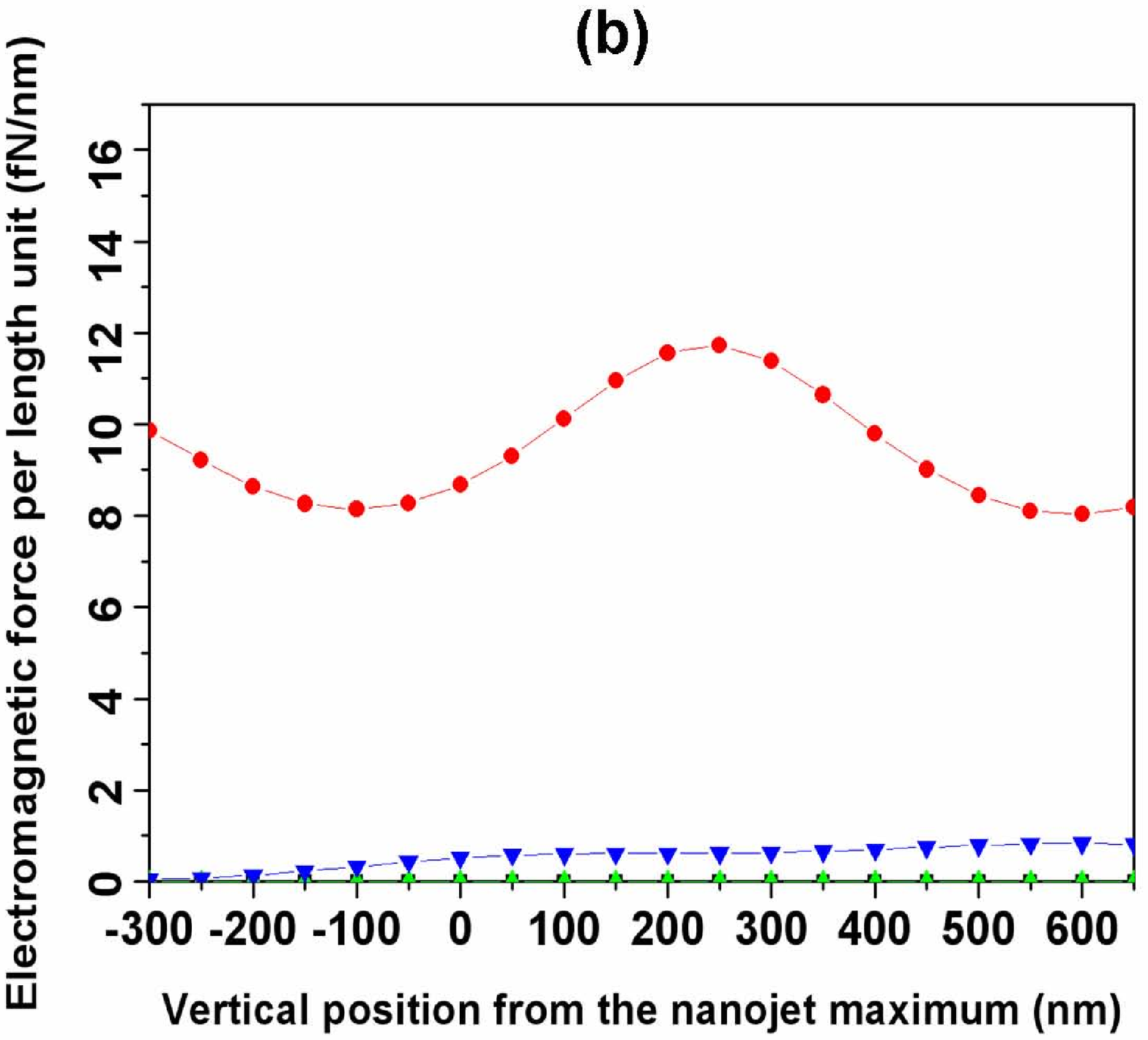}
\end{minipage}
\caption{(a) X- and Y-components of the total electromagnetic force ${\bf F}^t$ on the {\it Si} cylinder as it laterally moves to the right from the point of maximum intensity of the nanojet focused by the $SiO_{2}-glass$ cylinder of radius $r_{nj}=3\mu m$. This point is at $350 nm$ from the back surface of this $SiO_{2}-glass$ cylinder. No slit is present.  (b) The same quantities when the {\it Si} cylinder separates vertically from the same point as in (a). In both graphics,  black squares and green up-triangles  curves stand for ${\bf F}_{x}^{t}$ when the {\it Si} cylinder is non-resonant ($\lambda = 1280nm$) and when it is resonant ($\lambda = 1175nm$), respectively. Red circle and blue down-triangles represent, in the same order, ${\bf F}_{y}^t$.}
\end{figure}
Figure 5(a) shows the variation in the X- and Y-components of this force when the particle moves horizontally from the peak  point of the nanojet intensity map. In this case we assume no slit  present. The particle, illuminated out of resonance, is gradually and increasingly  attracted along the lateral direction by the gradient of intensity of the nanojet spatial distribution, and less and less pushed vertically, (cf. black squared and red circled curves, respectively). On the other hand, when the $WGH_{2,1}$ is excited, the particle is now slightly more and more laterally repelled from the peak intensity point of the nanojet map, (see the green up-triangle curve) and weakly, although increasingly and progressively, pushed in the vertical direction (cf. blue-down triangle curve).

Figure 5(b) presents the same quantities as those plotted in Fig. 5(a), but with the $Si$ cylinder vertically moving upwards from the maximum intensity point of the nanojet spatial distribution. The absence or presence of the resonance in this particle is shown by the black square and green up-triangle curves, respectively. Obviously, due to the symmetry of the field intensity map around this axis, no horizontal force is exerted.  In the vertical direction the particle is nevertheless  fairly strongly pushed upwards, the force magnitude following an oscillatory behavior due to the interference pattern between the field exiting the slit and that reflected down by the $Si$ cylinder. This push is weak if the WGM  resonance is excited. In this latter case, under  this $p$-polarization illumination most of the near field is concentrated inside the particle, with little intensity outside. This leads to such  weak optical forces on the resonant particle with the $p$-polarized WGM in both Figs. 5(a) and 5(b).

It is worth remarking that these forces ${\bf F}_{x}^{t}$ and ${\bf F}_{y}^{t}$ induced by the nanojet on the $Si$ particle are about  1000 and 10 times larger, respectively,  than those from direct illumination of this probe cylinder by the Gaussian beam like in a conventional optical tweezer, (compare the above Figs. 5(a) and 5(b) with Figs. 5(a) and 5(b) of \cite{ValdiviaValero2012_6}).

\begin{figure}[htbp]
\begin{minipage}{.49\linewidth}
\centering
\includegraphics[width=6cm]{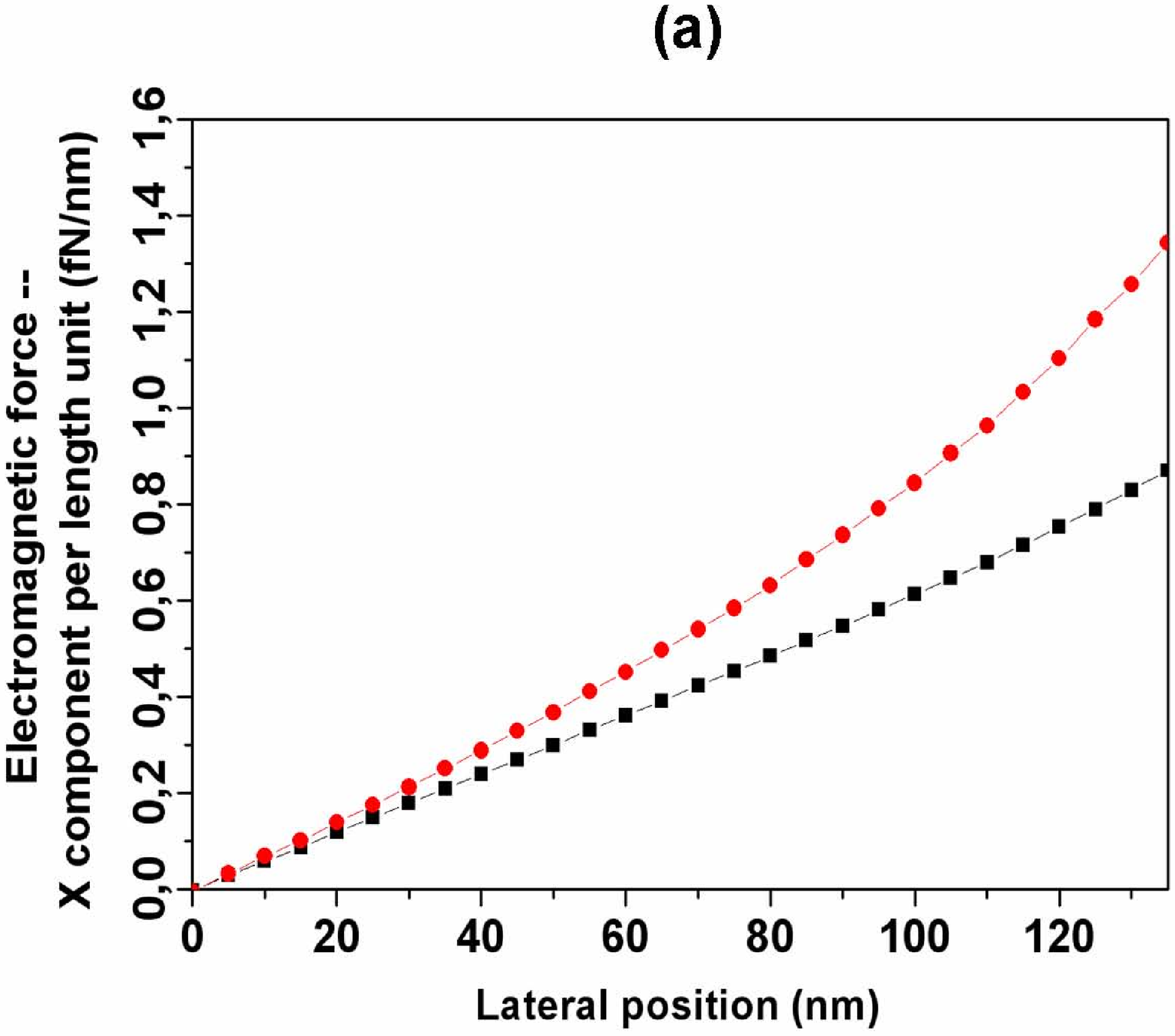}
\end{minipage}
\begin{minipage}{.49\linewidth}
\centering
\includegraphics[width=6cm]{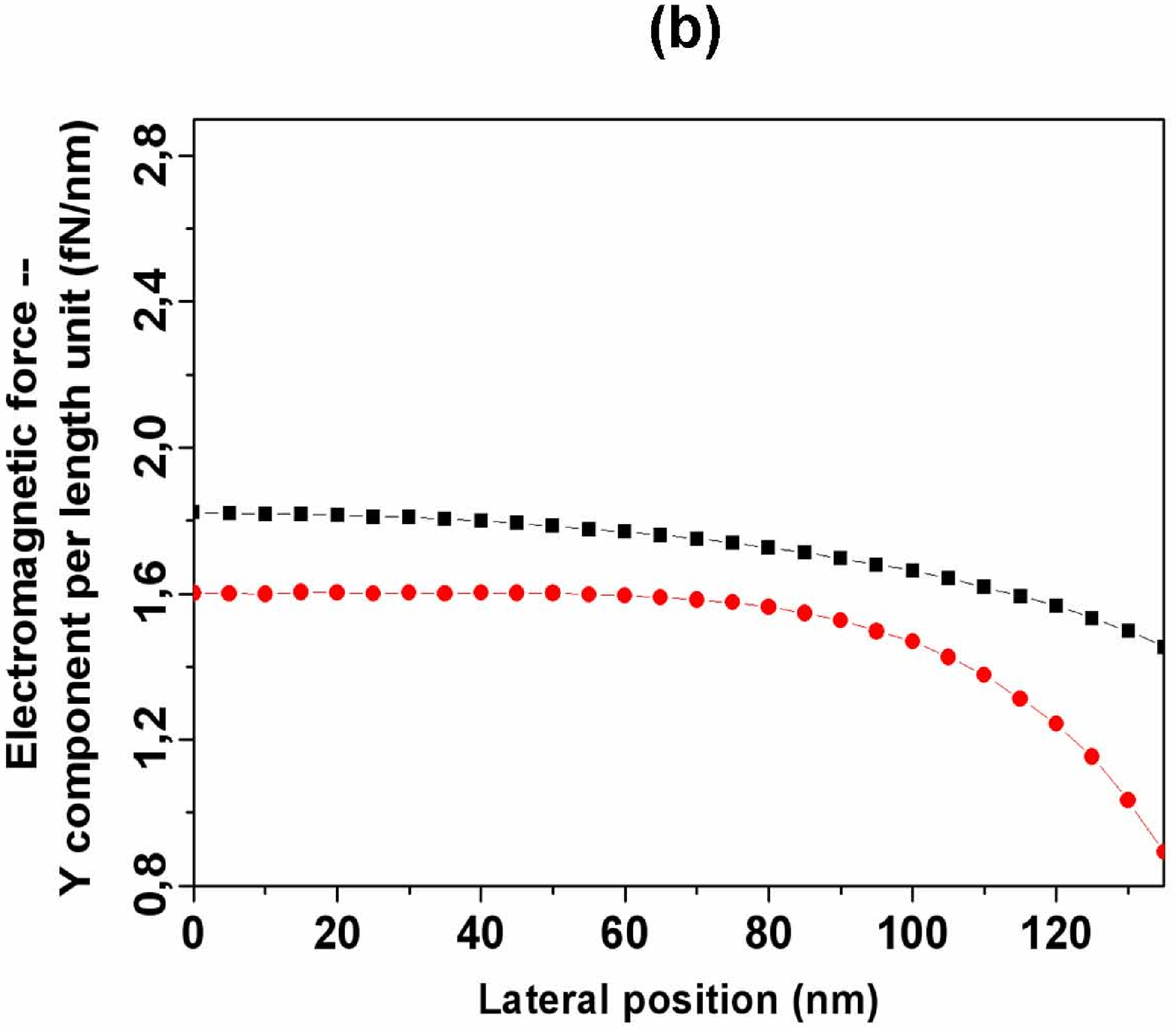}
\end{minipage}
\caption{(a) X-component ${\bf F}_{x}^t$ of the total electromagnetic force  exerted on the {\it Si} cylinder placed on the exit of the slit  practiced in the {\it Al} slab illuminated by a nanojet that arises near the back surface of the $SiO_{2}-glass$ cylinder of radius $r_{nj}=3\mu m$ below the slit, See Fig.1. The {\it Si} cylinder is tangent to the exit plane of the slit, and  laterally moves to the right from the aperture  axis. (b) Y-component  ${\bf F}_{y}^t$ of the optical force when the {\it Si} cylinder moves as in (a). In both graphics, black square and red circle curves stand for the cases in which the probe cylinder is non-resonant ($\lambda = 1280nm$) and  resonant ($\lambda = 1160nm$), respectively.}
\end{figure}

On the other hand, Figs. 6(a) and 6(b) show the X- and Y-components of the optical force exerted on the $Si$ cylinder,  placed  on the exit of the subwavelength slit, by the electromagnetic field exiting   it.  A $SiO_{2}-glass$ cylinder of radius $r_{nj}=3\mu m$ is below the slit and forms a nanojet near its back surface.  The particle laterally moves to the right approaching the slit edge. Each force Cartesian component is plotted for  the $Si$ cylinder being either out or in resonance. Now this particle is more and more horizontally attracted by the corners of the slit exit as it approaches them, independently of the absence or presence of the resonance [see the black square and red circle curves, respectively,  in Fig. 6(a)], the horizontal force being stronger in resonance.  Notwithstanding, the slit exerts a pushing vertical force on the cylinder which  decreases as it moves to the slit corners,  [cf. black squared and red circled curves, respectively, in Fig. 6(b)]. Then, the presence of the resonant $WGM_{2,1}$ in the $Si$ particle causes the vertical push from the slit to decrease and the horizontal attraction to the slit corners to increase. The cause of this is the same as discussed in connection with Figs. 5(a) and 5(b) for $p$-polarized illumination. However, as regards the horizontal force near the slit edge, the gradient force created around this edge by its concentration of charge, [see Figs. 4(b) and 4(c)], is  responsible for the larger value of ${\bf F}_{x}^t$ in resonance near this edge, as seen in Fig. 6(a). Thus, solely the field intensity gradient near the exit corners of the slit, (remember that in resonance under p-polarization the $Si$ cylinder concentrates the intensity inside, leaving very little scattered energy in its surroundings), is the major factor concerning the interaction between the aperture and the resonant particle.

These forces induced by the nanojet on the $Si$ particle through the slit are about  3 and 6 times larger   than those from direct illumination of the slit   by the Gaussian beam, i.e. without nanojet, (compare the above Figs. 6(a) and 6(b) with Figs. 5(a) and 5(b) of \cite{ValdiviaValero2012_6}).

\section{Extraordinary transmission of a subwavelength  slit illuminated by a nanojet. Excitation of  a localized surface plasmon.}

\subsection{Excitation of a  localized surface plasmon in a cylinder by nanojet focalization.  Effects of coupling by supertransmission}

We now deal with a subwavelength slit illuminated by a nanojet in the ultraviolet region when a metallic cylinder is at its exit. This produces two interacting processes of extraordinary transmission, or supertransmission,  enhancement: one due to the coupling of  the  nanojet focused in the aperture with the excitation of its propagating eigenmodes, and that arising from the coupling of the field transmitted through the slit  with the  localized surface plasmon (LSP) excited on the metallic cylinder,  \cite{ValdiviaValero2011_3, ValdiviaValero2012_6}.  In order to analyze the optical force on the metallic particle, we first briefly discuss the distribution of fields and energy in this system.

\begin{figure}[htbp]
\begin{minipage}{.49\linewidth}
\centering
\includegraphics[width=6cm]{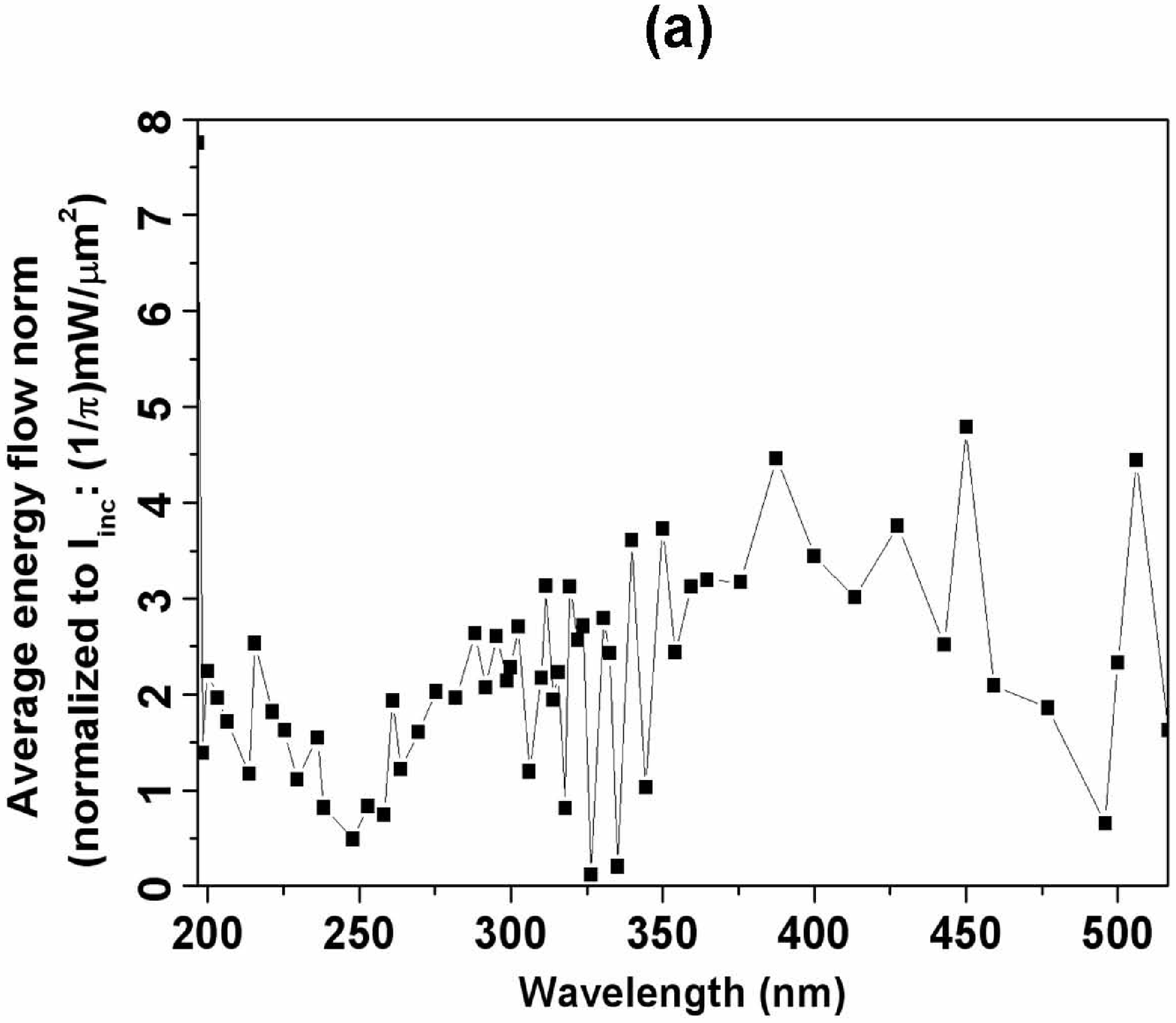}
\end{minipage}
\begin{minipage}{.49\linewidth}
\centering
\includegraphics[width=6cm]{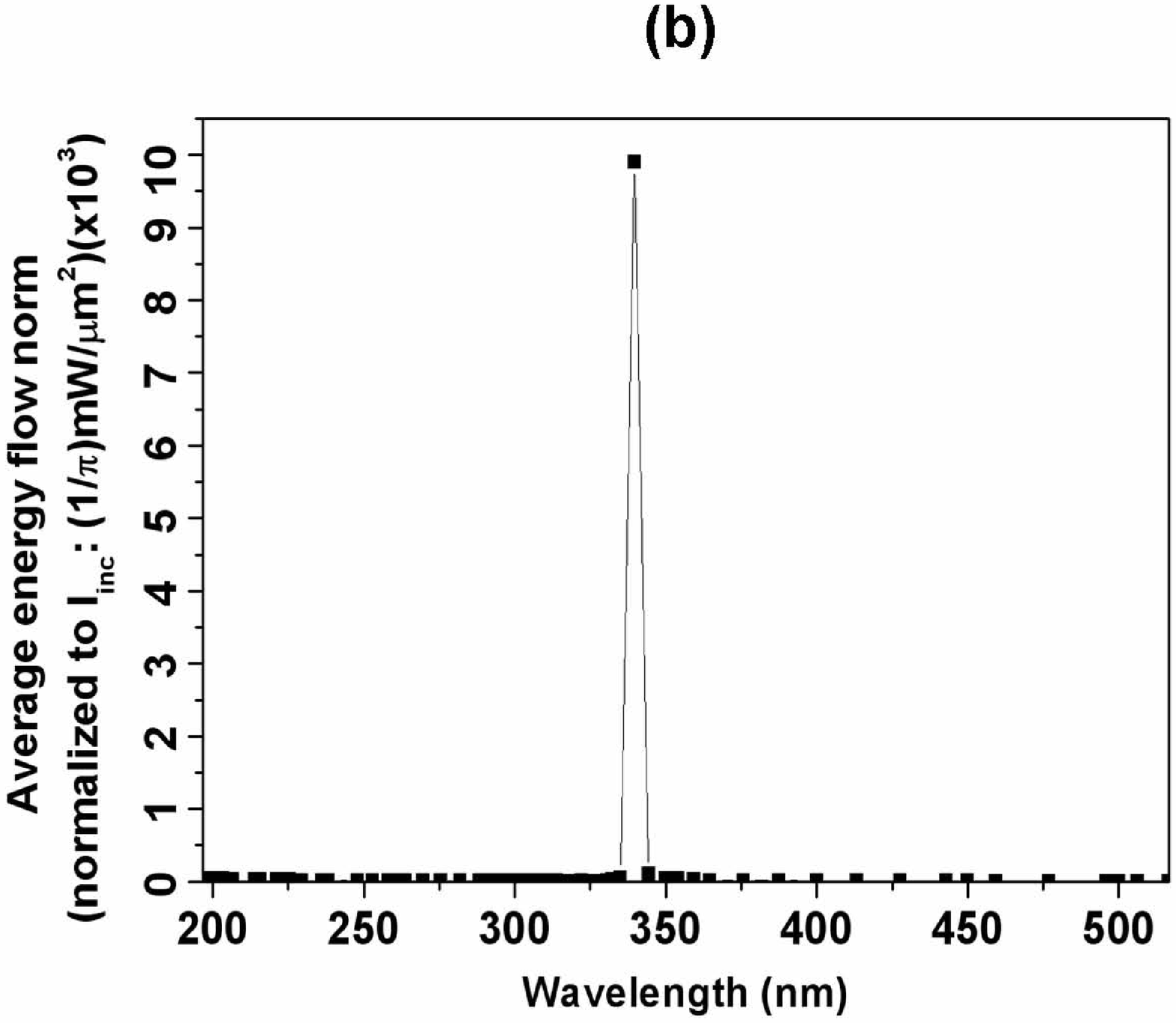}
\end{minipage}
\begin{minipage}{.98\linewidth}
\centering
\includegraphics[width=6cm]{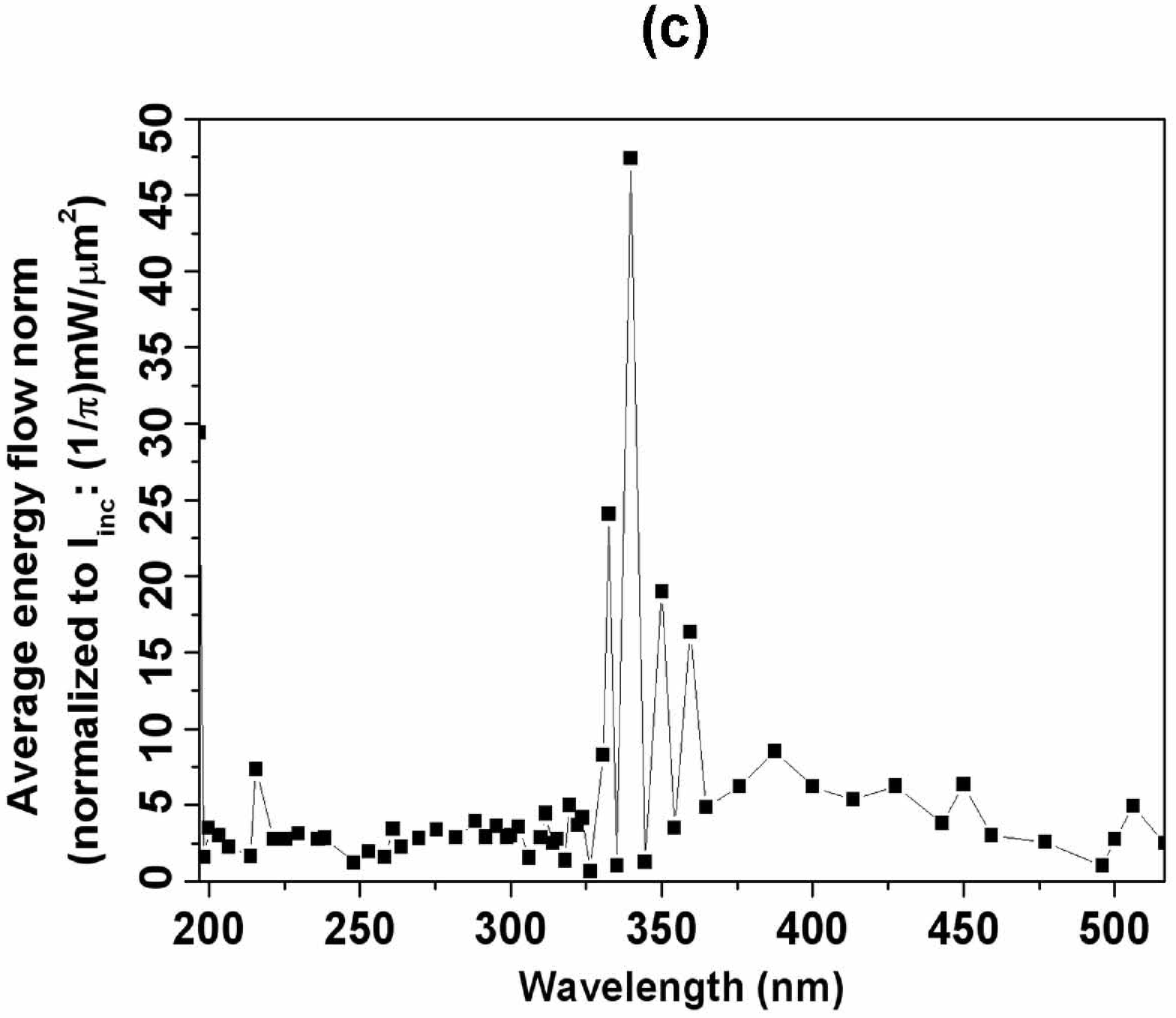}
\end{minipage}
\caption{(a) Time-averaged energy flow norm ${\bf |<S(r)>|}$ versus illumination wavelength, transmitted at the exit of a slit (width $d = 109.7nm$) practiced in an {\it Al} slab (width $D = 5.096\mu m$, thickness $h = 219.4nm$) illuminated by a nanojet. This nanojet is focused by a $SiO_{2}-glass$ cylinder of radius $r_{nj} = 1.3\mu m$, {\it 10nm} separated from the entrance plane of the slit. (b) No  slit is present: ${\bf |<S(r)>|}$  against illumination wavelength, concentrated on the surface of an {\it Ag} cylinder of radius $r = 30nm$ placed at {\it 200nm} from the back surface of the focusing  $SiO_{2}-glass$ cylinder. (c) The same quantity when the slit is present and the {\it Ag} cylinder, tangent to the exit plane of the aperture, scatters the light exiting the slit which is illuminated by the nanojet. These  calculations are done as  explained in Fig. 1.}
\end{figure}

Now a $SiO_{2}-glass$ cylinder of radius $r_{nj} = 1.3\mu m$ is placed below an $Al$ slab and focuses, near its back surface, a nanojet which illuminates the slit. The calculations are done as described in  Fig. 1. The rather smooth transmitted intensity   with peak at $\lambda = 339.7nm$ that would be obtained if this slit  were illuminated by the Gaussian beam without any nanojet, normalized to the maximum intensity (which was $0.75mW/\mu m^2$), becomes, when the nanojet illuminates the slit, an oscillatory line, typical of nanojets, as seen in Fig. 7(a).  Nevertheless, this configuration may render a slit-nanojet transmittance enhancement factor  of  about 4. Figure 7(b) represents the effect of light concentration in an $Ag$ cylinder of radius $r = 30nm$ placed at {\it 200nm} from the back surface of the focusing $SiO_{2}-glass$ cylinder when the slit is suppressed. The excited $TM_{2,1}$  eigenmode energy concentrated on the  surface of the $Ag$ cylinder is enhanced  at $\lambda = 339.7nm$ by a factor about 33 due to the presence of the nanojet, [see Fig. 7(b)]. This effect is somewhat similar to that on the above studied $Si$ particle, although now with a much stronger enhancement factor.

On the other hand, when the slit is also present as shown in Fig. 7(c), enhancement factors between 8 and 47 are obtained on ecitation of the $TM_{2,1}$ LSP in the $Ag$ cylinder.
\begin{figure}[htbp]
\begin{minipage}{.49\linewidth}
\centering
\includegraphics[width=6cm]{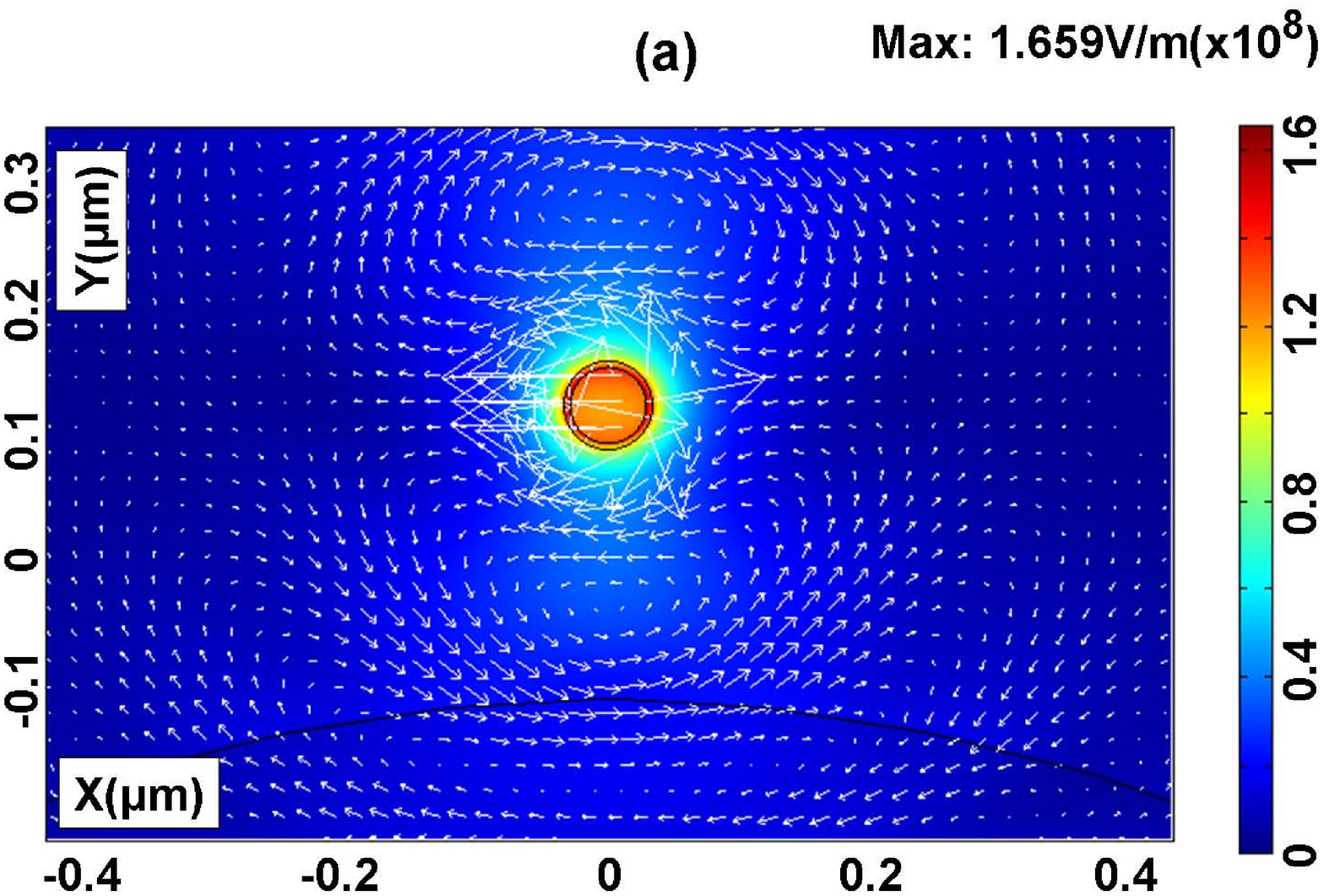}
\end{minipage}
\begin{minipage}{.49\linewidth}
\centering
\includegraphics[width=6cm]{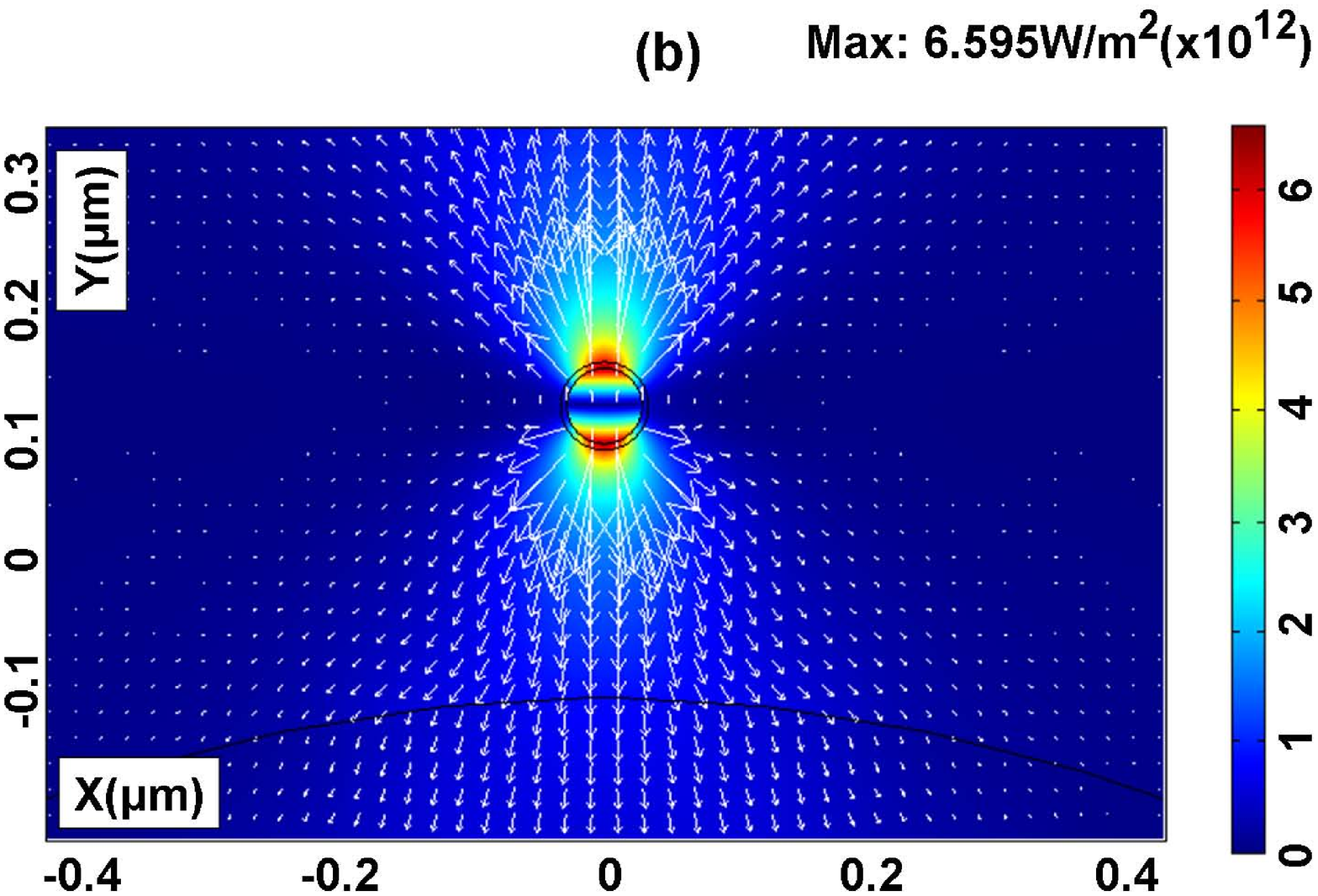}
\end{minipage}
\begin{minipage}{.49\linewidth}
\centering
\includegraphics[width=6cm]{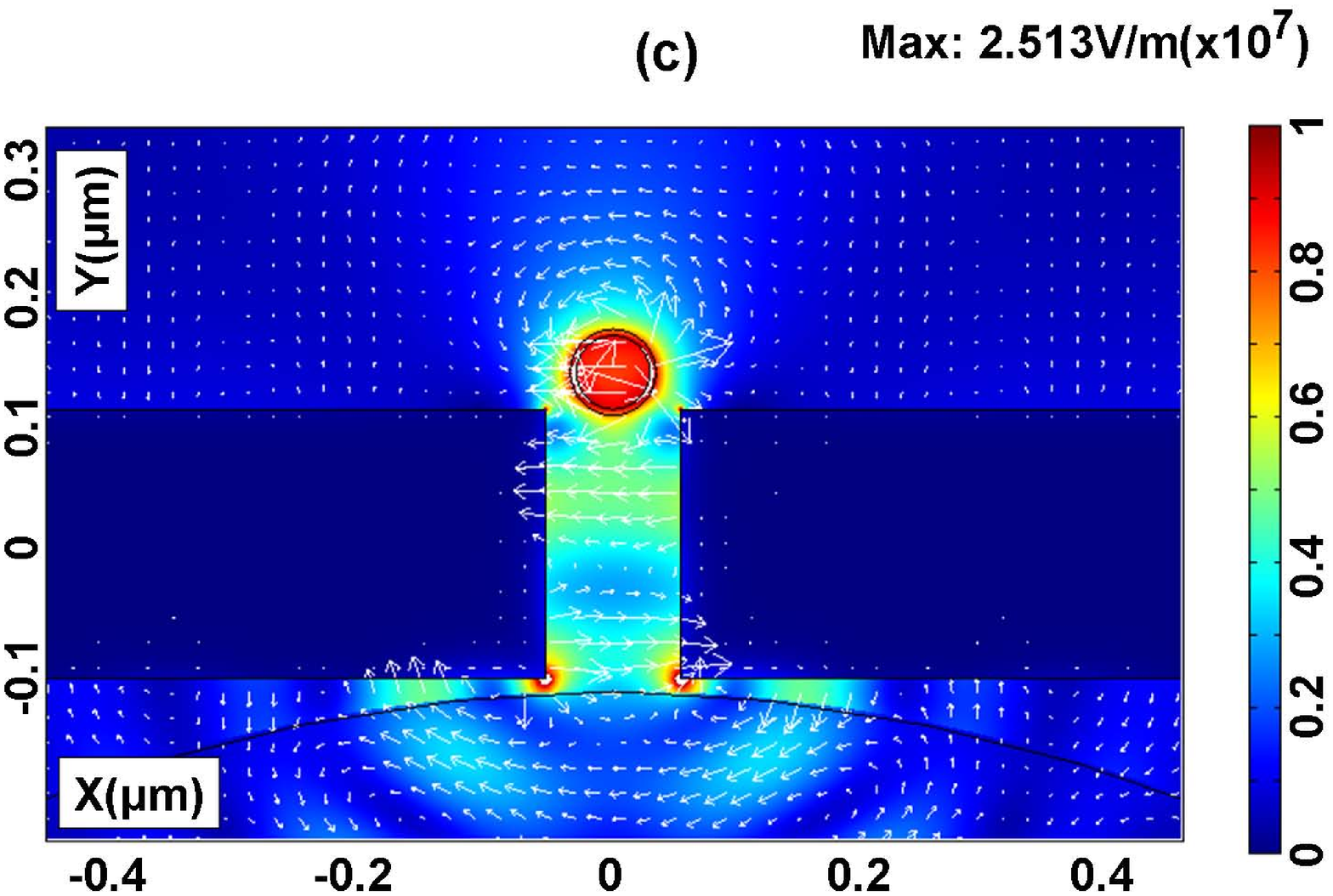}
\end{minipage}
\begin{minipage}{.49\linewidth}
\centering
\includegraphics[width=6cm]{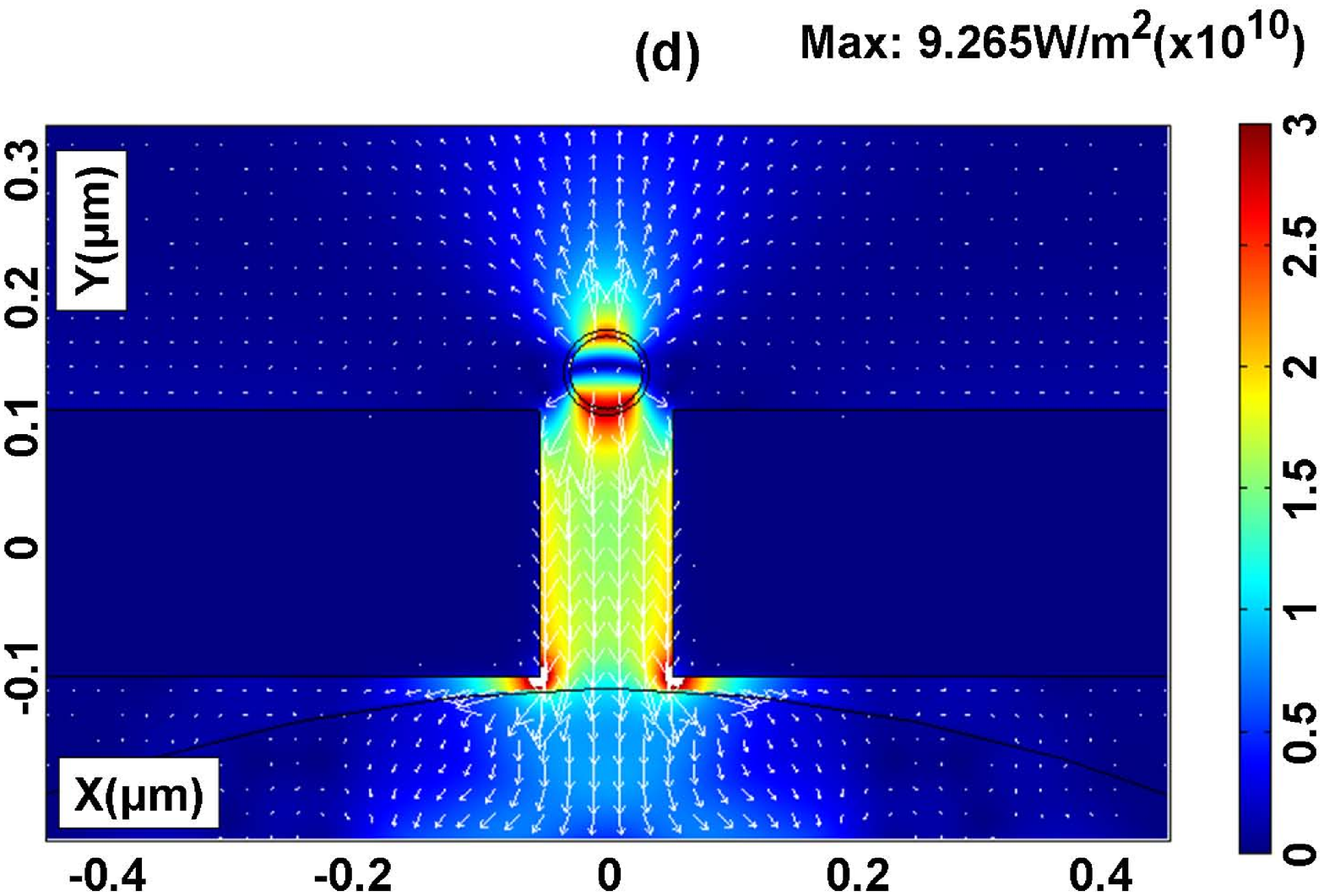}
\end{minipage}
\caption{(a) Detail of the electric field spatial distribution ${\bf E(r)}$, ${\bf r}=(x,y)$, (magnitude and vector:  colors and arrows, respectively), near the $Ag$ cylinder of radius $r = 30nm$ placed at {\it 200nm} from the back surface of the focusing  $SiO_{2}-glass$ cylinder of radius $r_{nj} = 1.3\mu m$. No slit is present. (b) The same detail  but now in time-averaged energy flow ${\bf <S(r)>}$, (norm in colors and vector in arrows). (c) Detail of ${\bf E(r)}$ (magnitude and vector:  colors and arrows, respectively), when the slit is also present. The configuration corresponds to the largest peak in of Fig. 7(c). ${\bf |E(r)|}_{max} = 2.513\cdot 10^7V/m$. (d) The same detail as in (c),  now in ${\bf <S(r)>}$ (norm in colors and vector in arrows). ${\bf |<S(r)>|}_{max} = 9.265\cdot 10^{10}W/m^2$. In all images the illumination wavelength is $\lambda = 339.7nm$.}
\end{figure}

As illustrations, Figs. 8(a) and 8(b) present details of the effect of the nanojet near  the $Ag$ cylinder. As seen, there is a skin depth effect at the illuminating wavelength due to the small size of this  $Ag$ cylinder. This case  corresponds to the largest peak  in Fig. 7(b). The electric field ${\bf E(r)}$ and the time-averaged energy flow ${\bf <S(r)>}$, respectively, are also shown. The nanojet is almost completely concentrated in the small metallic cylinder. Figs. 8(c) and 8(d) are details of ${\bf E(r)}$ and ${\bf <S(r)>}$, respectively, for the configuration studied in Fig. 7(c) i. e. that of the nanojet-LSP coupling through the slit. The wavelength $\lambda = 339.7nm$ corresponds to the largest peak in Fig. 7(c),  which practically remains the same as that of Fig. 7(b) when there is no slit.  The hotspots in the slit edge, the  nanojet focalization as well as the LSP excitation mechanisms are shown.

\subsection{Electromagnetic forces on a metallic cylinder with a localized surface plasmon excited by a nanojet,  either directly or on transmission  through a subwavelength slit.}

We now are in position to analyze the  optical forces induced by the nanojet and the slit on the metallic $Ag$ particle with excitation $LSPs$.

\begin{figure}[htbp]
\begin{minipage}{.49\linewidth}
\centering
\includegraphics[width=6cm]{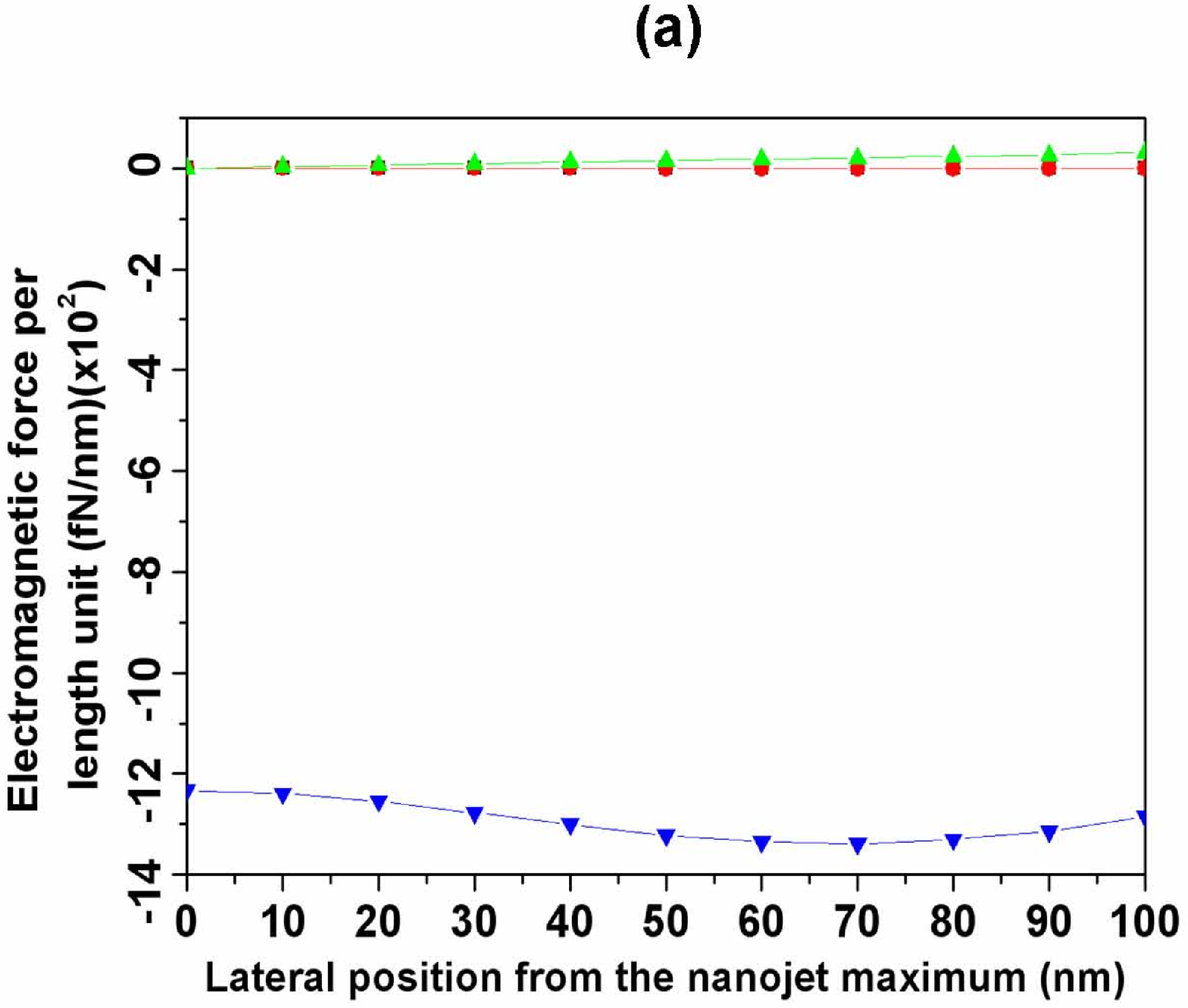}
\end{minipage}
\begin{minipage}{.49\linewidth}
\centering
\includegraphics[width=6cm]{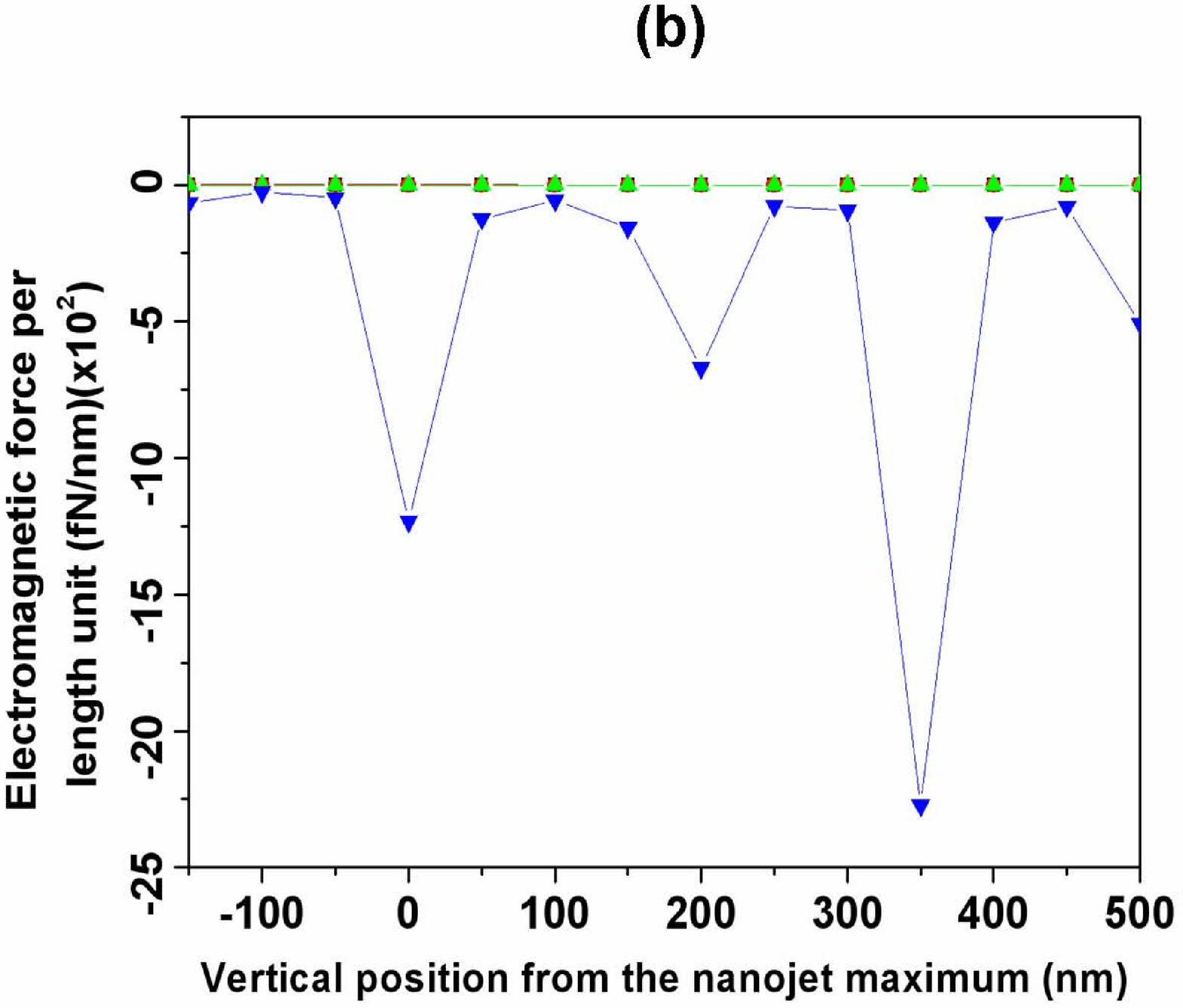}
\end{minipage}
\caption{(a) X- and Y-components of the total electromagnetic force ${\bf F}^t$, calculated as explained in Fig. 1 and Eq. (1), due to the nanojet focused by the $SiO_{2}-glass$ cylinder, acting on the {\it Ag} cylinder of radius $r = 30nm$ as this laterally moves  to the right  from the point of maximum intensity of the nanojet. (This point is on the vertical symmetry axis at 200nm from the back surface of the $SiO_{2}-glass$ cylinder). No slit is present. (b) The same when the {\it Ag} cylinder  vertically moves upwards from the point of maximum intensity of the nanojet.  In both figures, black squared and triangle up green curves stand for ${\bf F}_{x}^t$, both when the {\it Ag} cylinder is non-resonant ($\lambda = 500.0nm$) and when it is resonant ($\lambda = 339.7nm$), respectively. Red circle and triangle down blue lines represent ${\bf F}_{y}^t$ in the same order and  for the same cases as before.}
\end{figure}

To this end, the excitation of  the $Ag$ cylinder $TM_{2,1}$ LSP eigenmode by focusing a nanojet on it, is studied. Figure 9(a) shows the behavior of the X- and Y-components of the force on this metallic particle as it progressively moves to the right in absence of the slit. This force is due to only the field of the nanojet scattered by the $Ag$ particle.  When the particle is not resonant, the forces along the X and Y directions are very weak, of the order of $10^{-2}fN/nm$ and $10^{-1}fN/nm$ for the horizontal force (black curve with squares) and for the vertical force (red curve with circles), respectively. In resonance the horizontal force is still very weak, but the vertical component is strong and atractive. On the other hand, as the $Ag$ cylinder moves upwards along the nanojet axis, as shown in Fig. 9(b), no transversal force appears (black squared and green up-triangled curves, respectively) as it should, but  differences in the vertical force arise depending on the absence or the presence of a LSP resonance. This vertical force is very small in absence of LSP excitation: about $10^{-1}fN/nm$, whereas it is attractive and oscillating with distance, and of the order of $10^3fN/nm$ when the LSP resonance of the $Ag$ cylinder is excited, this oscillation is due to the interference pattern between the nanojet inciding on the metallic particle and the field that the latter scatters backward.

Notice the remarkable pulling nature of this vertical force, which converts the  $SiO_{2}-glass$ cylinder in a photonic tractor \cite{JuanjoNat, Chinos, Novitski} for certain plasmonic small particles. This contrasts with the repulsive  ${\bf F}_{y}^t$ obtained in \cite{Cui2008} on an $Ag$ particle in resonance. However, if one doubles its  radius making it of $60 nm$, then this force becomes repulsive due to its larger scattering cross section. We do not show this here for brevity.

\begin{figure}[htbp]
\begin{minipage}{.49\linewidth}
\centering
\includegraphics[width=6cm]{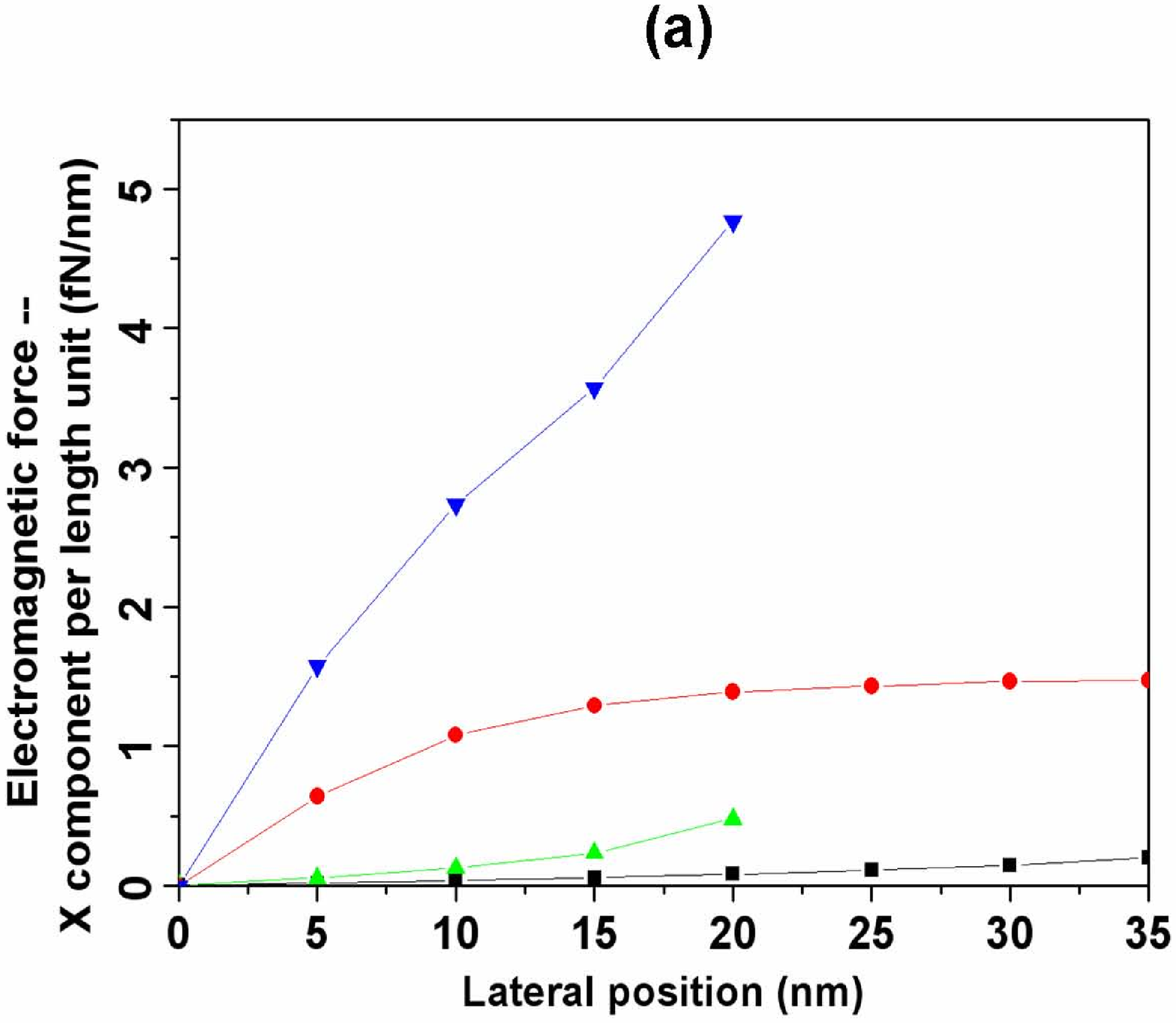}
\end{minipage}
\begin{minipage}{.49\linewidth}
\centering
\includegraphics[width=6cm]{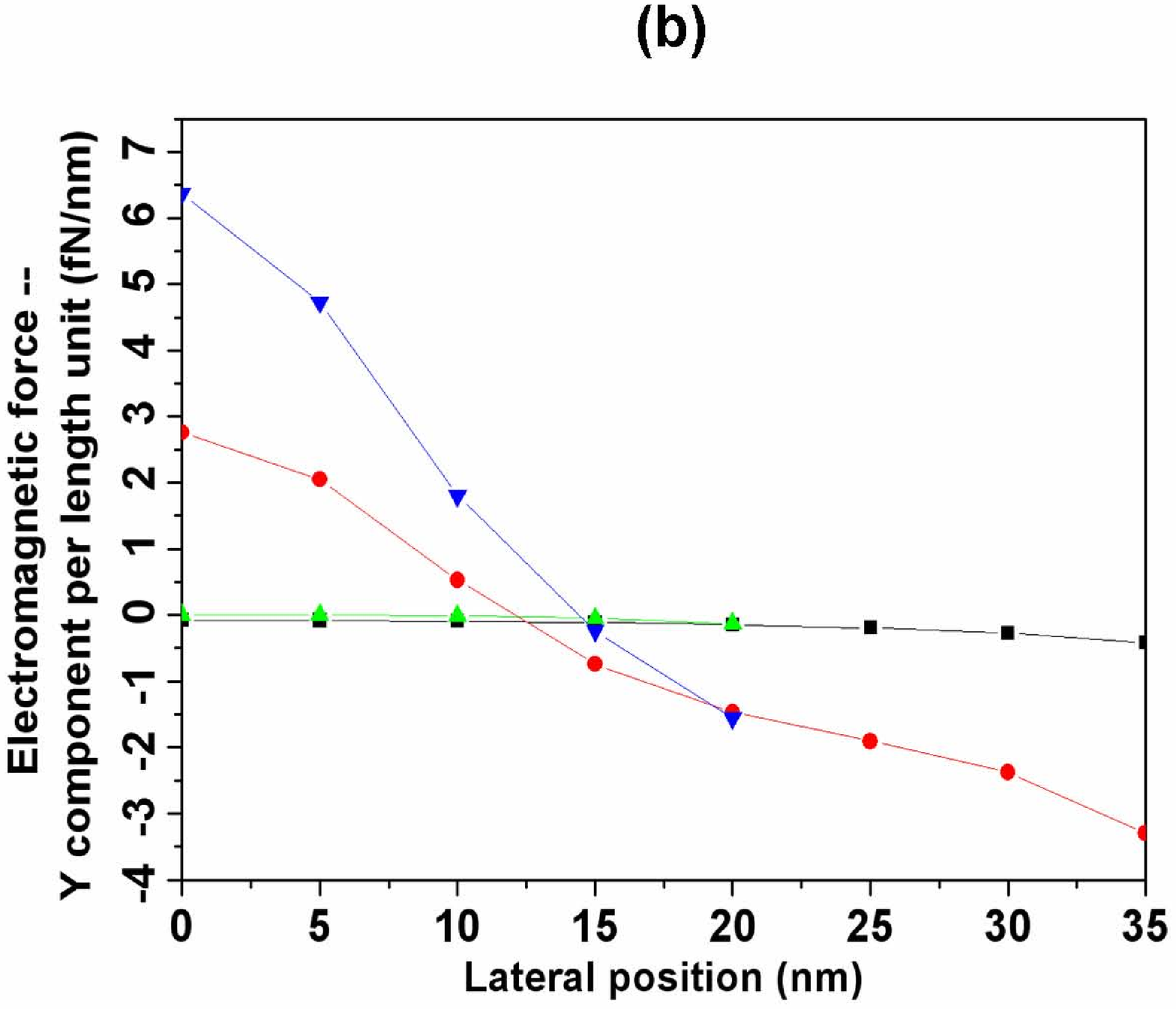}
\end{minipage}
\caption{(a) X-component ${\bf F}_{x}^t$ of the total electromagnetic force exerted on the {\it Ag} cylinder of radius $r = 30nm$, calculated as explained in Fig. 1 and Eq. (1). The slit is practiced in the {\it Al} slab, being illuminated by the nanojet which arises from the back surface of the $SiO_{2}-glass$ cylinder placed below. The {\it Ag} cylinder is tangent to the exit plane of the slit, and laterally moves to the right from the aperture axis. (b) Y-component ${\bf F}_{y}^t$ of the optical force when the {\it Ag} cylinder moves as in (a). In both graphics, black square and red circle curves stand for the cases in which the {\it Ag} cylinder, tangent to the exit line of the aperture, is non-resonant ($\lambda = 1280nm$) and resonant ($\lambda = 1160nm$), respectively. The green up triangles and blue down triangles curves represent, in the same order as before, the same cases when the {\it Ag} cylinder is at the exit slit, {\it 20nm} inwards.}
\end{figure}

Finally, the presence of the slit  illuminated by the nanojet is analized concerning the optical forces acting on the $Ag$ cylinder. Once again, the excitation of a LSP resonance largely affects these forces, but due to $p$-polarized illumination, it has consequences opposite to those of the WGM excitation in a  dielectric particle, (cf. Figs. 10(a) and 10(b) for the  X- and Y-components of this optical force, respectively). Black square and red circle curves in Fig. 10(a) stand for the cases of absence and presence of LSP resonance in the $Ag$ cylinder, respectively. They show that the plasmonic resonance enhances the lateral force, which progressively attracts the particle towards the corners as it approaches it in its displacement. The vertical force, plotted with the same color code in Fig. 10(b), behaves in a more complex way. This being slightly attractive when the $Ag$ cylinder is non resonant (values of order $10^{-1}fN/nm$). However, in presence of resonance this vertical force is, though, repulsive in the central region of the slit exit, becoming attractive near its corners (red circled curve). These tendencies can be increased by locating the $Ag$ cylinder inwards the slit exit, as noticed by the green up-triangles and blue down-triangles curves of both Figs. 10(a) and 10(b). [Values of order $10^{-1}fN/nm$ for the green up triangles curve (non resonant cylinder) in Fig. 10(b)].

In any case, the  forces induced by the nanojet on the $Ag$ particle through the slit are about  10 times larger   than those from direct illumination of the slit   by the Gaussian beam without nanojet, (compare the above Figs. 10(a) and 10(b) with Figs. 11(a) and 11(b) of \cite{ValdiviaValero2012_6}).

\section{Discussion and conclusions}
We have made a study on the photonic forces on nanoparticles near a subwavelength slit illuminated by a nanojet.  We have performed comparisons of the  force on the particles either in or out of their Mie resonance excitation.  The study has been done in 2D, so that these objects are cylinders and the transmission modes of the slit are excited by  p-polarized illumination.   However, we believe that these results also hold for spheres in 3D, aside of polarization effects.

We have proven that whereas the morphological resonance of the slit, causing supertransmission, enhances the fields surrounding the probe cylinders and hence the optical forces exerted upon them, the excitation of the probe particle resonance has a different effect on these forces, according to whether the cylinder is dielectric or metallic.

The photonic force on the probe particle  dependens on the field surrounding it. Thus  WGMs  in dielectric nanocylinders under p-polarization, (which is the one at which supertransmission exists in 2D), have no much effects on the optical force. Its strength  may weaken due to the low  intensity distribution  in the near field region of this particle. In this connection, it must be remarked that the electromagnetic force on dielectric cylinders is however enhanced under s-polarization illumination \cite{AriasGonzalez2002}; then the WGM extends to the near field outside the cylinder.

On the other hand, LSPs in metallic nanocylinders give rise to localized field energy enhancement on the cylinder surface; this strengthes the optical force. In contrast with illumination by only a PNJ, the presence of a subwavelength slit enriches the effect of the electromagnetic forces. Whereas such a a small metallic particle resonantly illuminated by a nanojet, suffers an attractive vertical force  which makes the
 $SiO_{2}-glass$ nanojet forming cylinder to behave as  a photonic tractor,  the additional presence of a subwavelength aperture may change the sign of this force thus transforming it to repulsive, depending on the lateral position of the cylinder.

 For any cylinder, whether dielectric or metallic, we observe that illuminating the slit with a  nanojet increases the magnitude of the optical forces on the particle by a factor between 3 and 10 compared with illumination of the slit with a Gausian beam. On the other hand, the gradient horizontal force induced by a nanojet in a dielectric particle is a few orders of magnitude larger than that of a Gaussian beam of a conventional optical tweezer.

These results should stimulate experiments with PNJs on slits since they are the result of focusing by microparticles, and their highly localized intensity  at the subwavelength scale present an ideal means to spatially control the mechanical action on nanoobjects.

%After proofreading the manuscript, tar and gzip the \texttt{.tex} file and
%figures; then enter the requested information into the \textit{Optics Express}
%online submission system at \url{http://www.opticsexpress.org} and upload the
%tarred and gzipped archive. If there is video or other multimedia, the associated
%files should be uploaded separately.

\section*{Acknowledgements}

Work supported by the Spanish MEC through FIS2009-13430-C02-C01 and Consolider NanoLight (CSD2007-00046) research grants, FJVV is supported by the last grant.

\end{document}